\documentclass[preprint]{elsarticle}
\pdfoutput=1
\usepackage{ifthen}
\newboolean{uprightparticles}
\setboolean{uprightparticles}{false} %Set true for upright particle symbols

\usepackage{amssymb,epsf,epsfig,amsmath,color, array}
\usepackage[colorlinks,allcolors=blue]{hyperref}
\usepackage{lscape}
\usepackage{mciteplus}
\usepackage{ifthen}
\usepackage{xspace}
\usepackage{graphicx}
\usepackage{amssymb}
\usepackage{amsmath}
\usepackage{xspace}
\usepackage{float}
\usepackage{lscape}
\usepackage{caption}
\usepackage{subfigure}
\usepackage{multirow}
\usepackage{cleveref}
\usepackage{rotating}
\usepackage[T1]{fontenc}
\usepackage{footnote}
\usepackage{tablefootnote}
\makesavenoteenv{sidewaystable}
\makesavenoteenv{tabular}
\usepackage{etoolbox}
\AtBeginEnvironment{tabular}{\small\setlength{\tabcolsep}{1pt}}
\usepackage[para,flushleft]{threeparttable}
\usepackage{upgreek}

\def\lhcb {\mbox{LHCb}\xspace}

\def\babar  {\mbox{BaBar}\xspace}
\def\belle  {\mbox{Belle}\xspace}

\def\MagUp {\mbox{\em Mag\kern -0.05em Up}\xspace}

\ifthenelse{\boolean{uprightparticles}}%
{
 
 \def\Pgamma      {\ensuremath{\upgamma}\xspace}

 \def\Pmu         {\ensuremath{\upmu}\xspace}                 
 \def\Pnu         {\ensuremath{\upnu}\xspace}                 
                  
 \def\Ppi         {\ensuremath{\uppi}\xspace}

 \def\Ptau        {\ensuremath{\uptau}\xspace}

 \def\PDelta      {\ensuremath{\Delta}\xspace}                 
 \def\PXi      {\ensuremath{\Xi}\xspace}                 
 \def\PLambda      {\ensuremath{\Lambda}\xspace}                 
 \def\PSigma      {\ensuremath{\Sigma}\xspace}                 
 \def\POmega      {\ensuremath{\Omega}\xspace}                 
 \def\PUpsilon      {\ensuremath{\Upsilon}\xspace}                 
 
 \def\PB      {\ensuremath{\mathrm{B}}\xspace}                 
                  
 \def\PD      {\ensuremath{\mathrm{D}}\xspace}

 \def\PK      {\ensuremath{\mathrm{K}}\xspace}

 \def\PZ      {\ensuremath{\mathrm{Z}}\xspace}                 
                  
 \def\Pb      {\ensuremath{\mathrm{b}}\xspace}                 
 \def\Pc      {\ensuremath{\mathrm{c}}\xspace}                 
 \def\Pd      {\ensuremath{\mathrm{d}}\xspace}                 
 \def\Pe      {\ensuremath{\mathrm{e}}\xspace}

 \def\Pi      {\ensuremath{\mathrm{i}}\xspace}

 \def\Pp      {\ensuremath{\mathrm{p}}\xspace}

 \def\Ps      {\ensuremath{\mathrm{s}}\xspace}                 
 \def\Pt      {\ensuremath{\mathrm{t}}\xspace}                 
 \def\Pu      {\ensuremath{\mathrm{u}}\xspace}

}
{
 
 \def\Pgamma      {\ensuremath{\gamma}\xspace}

 \def\Pmu         {\ensuremath{\mu}\xspace}                 
 \def\Pnu         {\ensuremath{\nu}\xspace}                 
                  
 \def\Ppi         {\ensuremath{\pi}\xspace}

 \def\Ptau        {\ensuremath{\tau}\xspace}

 \mathchardef\PDelta="7101
 \mathchardef\PXi="7104
 \mathchardef\PLambda="7103
 \mathchardef\PSigma="7106
 \mathchardef\POmega="710A
 \mathchardef\PUpsilon="7107
                  
 \def\PB      {\ensuremath{B}\xspace}                 
                  
 \def\PD      {\ensuremath{D}\xspace}

 \def\PK      {\ensuremath{K}\xspace}

 \def\PZ      {\ensuremath{Z}\xspace}                 
                  
 \def\Pb      {\ensuremath{b}\xspace}                 
 \def\Pc      {\ensuremath{c}\xspace}                 
 \def\Pd      {\ensuremath{d}\xspace}                 
 \def\Pe      {\ensuremath{e}\xspace}

 \def\Pi      {\ensuremath{i}\xspace}

 \def\Pp      {\ensuremath{p}\xspace}

 \def\Ps      {\ensuremath{s}\xspace}                 
 \def\Pt      {\ensuremath{t}\xspace}                 
 \def\Pu      {\ensuremath{u}\xspace}

}

\makeatother

\DeclareRobustCommand{\optbar}[1]{\shortstack{{\miniscule (\rule[.5ex]{1.25em}{.18mm})}
  \\ [-.7ex] $#1$}}

\def\electron   {{\ensuremath{\Pe}}\xspace}
\def\en         {{\ensuremath{\Pe^-}}\xspace}   % electron negative (\em is taken)
\def\ep         {{\ensuremath{\Pe^+}}\xspace}
\def\epm        {{\ensuremath{\Pe^\pm}}\xspace}

\def\muon       {{\ensuremath{\Pmu}}\xspace}
\def\mup        {{\ensuremath{\Pmu^+}}\xspace}
\def\mun        {{\ensuremath{\Pmu^-}}\xspace} % muon negative (\mum is taken)

\def\mump       {{\ensuremath{\Pmu^{\mp}}}\xspace}

\def\tauon      {{\ensuremath{\Ptau}}\xspace}
\def\taup       {{\ensuremath{\Ptau^+}}\xspace}
\def\taum       {{\ensuremath{\Ptau^-}}\xspace}

\def\ellm       {{\ensuremath{\ell^-}}\xspace}
\def\ellp       {{\ensuremath{\ell^+}}\xspace}
\def\ellell     {\ensuremath{\ell^+ \ell^-}\xspace}

\def\neu        {{\ensuremath{\Pnu}}\xspace}
\def\neub       {{\ensuremath{\overline{\Pnu}}}\xspace}

\def\neumb      {{\ensuremath{\neub_\mu}}\xspace}
\def\neut       {{\ensuremath{\neu_\tau}}\xspace}
\def\neutb      {{\ensuremath{\neub_\tau}}\xspace}

\def\neulb      {{\ensuremath{\neub_\ell}}\xspace}
\def\g      {{\ensuremath{\Pgamma}}\xspace}

\def\Z      {{\ensuremath{\PZ}}\xspace}

\def\uquark    {{\ensuremath{\Pu}}\xspace}
\def\uquarkbar {{\ensuremath{\overline \uquark}}\xspace}

\def\dquark    {{\ensuremath{\Pd}}\xspace}

\def\squark    {{\ensuremath{\Ps}}\xspace}

\def\cquark    {{\ensuremath{\Pc}}\xspace}

\def\bquark    {{\ensuremath{\Pb}}\xspace}
\def\bquarkbar {{\ensuremath{\overline \bquark}}\xspace}

\def\tquark    {{\ensuremath{\Pt}}\xspace}

\def\pion   {{\ensuremath{\Ppi}}\xspace}

\def\pip    {{\ensuremath{\pion^+}}\xspace}
\def\pim    {{\ensuremath{\pion^-}}\xspace}

\def\kaon    {{\ensuremath{\PK}}\xspace}
  \def\Kbar    {{\kern 0.2em\overline{\kern -0.2em \PK}{}}\xspace}

\def\KorKbar    {\kern 0.18em\optbar{\kern -0.18em K}{}\xspace}

\def\Kp      {{\ensuremath{\kaon^+}}\xspace}
\def\Km      {{\ensuremath{\kaon^-}}\xspace}

\def\Kstarz  {{\ensuremath{\kaon^{*0}}}\xspace}

\def\Kstar   {{\ensuremath{\kaon^*}}\xspace}

\def\Kstarp  {{\ensuremath{\kaon^{*+}}}\xspace}

  \def\Dbar    {{\kern 0.2em\overline{\kern -0.2em \PD}{}}\xspace}
\def\D       {{\ensuremath{\PD}}\xspace}

\def\DorDbar    {\kern 0.18em\optbar{\kern -0.18em D}{}\xspace}
\def\Dz      {{\ensuremath{\D^0}}\xspace}

\def\Dstar   {{\ensuremath{\D^*}}\xspace}

\def\Dstarp  {{\ensuremath{\D^{*+}}}\xspace}

\def\B       {{\ensuremath{\PB}}\xspace}
\def\Bbar    {{\ensuremath{\kern 0.18em\overline{\kern -0.18em \PB}{}}}\xspace}

\def\BorBbar    {\kern 0.18em\optbar{\kern -0.18em B}{}\xspace}
\def\Bz      {{\ensuremath{\B^0}}\xspace}

\def\Bu      {{\ensuremath{\B^+}}\xspace}
\def\Bub     {{\ensuremath{\B^-}}\xspace}
\def\Bp      {{\ensuremath{\Bu}}\xspace}
\def\Bm      {{\ensuremath{\Bub}}\xspace}

\def\Bd      {{\ensuremath{\B^0}}\xspace}
\def\Bs      {{\ensuremath{\B^0_\squark}}\xspace}

  \def\Y#1S{\ensuremath{\PUpsilon{(#1S)}}\xspace}% no space before {...}!

\def\proton      {{\ensuremath{\Pp}}\xspace}

\def\Lz          {{\ensuremath{\PLambda}}\xspace}
\def\Lbar        {{\ensuremath{\kern 0.1em\overline{\kern -0.1em\PLambda}}}\xspace}
\def\LorLbar    {\kern 0.18em\optbar{\kern -0.18em \PLambda}{}\xspace}

\def\Lb      {{\ensuremath{\Lz^0_\bquark}}\xspace}

\def\Lc      {{\ensuremath{\Lz^+_\cquark}}\xspace}

\def\BF         {{\ensuremath{\mathcal{B}}}\xspace}

\newcommand{\decay}[2]{\ensuremath{#1\!\to #2}\xspace}         % {\Pa}{\Pb \Pc}

\def\to                 {\ensuremath{\rightarrow}\xspace}

\def\qsq       {{\ensuremath{q^2}}\xspace}

\def\CP                {{\ensuremath{C\!P}}\xspace}

\def\rhobar {{\ensuremath{\overline \rho}}\xspace}
\def\etabar {{\ensuremath{\overline \eta}}\xspace}

\def\Vub  {{\ensuremath{\left| V_{\uquark\bquark} \right|}}\xspace}
\def\Vcb  {{\ensuremath{\left| V_{\cquark\bquark} \right|}}\xspace}
\def\Vtb  {{\ensuremath{V_{\tquark\bquark}}}\xspace}

\def\Vtss  {{\ensuremath{V_{\tquark\squark}^\ast}}\xspace}

\newcommand{\dms}{{\ensuremath{\Delta m_{\squark}}}\xspace}
\newcommand{\dmd}{{\ensuremath{\Delta m_{\dquark}}}\xspace}
\newcommand{\DG}{{\ensuremath{\Delta\Gamma}}\xspace}

\newcommand{\ACP}{{\ensuremath{{\mathcal{A}}^{\CP}}}\xspace}

\def\bsll     {\decay{\bquark}{\squark \ell^+ \ell^-}}

\def\AT#1     {\ensuremath{A_{\mathrm{T}}^{#1}}\xspace}           % 2

\def\C#1      {\ensuremath{\mathcal{C}_{#1}}\xspace}                       % 9
\def\Cp#1     {\ensuremath{\mathcal{C}_{#1}^{'}}\xspace}                    % 7
\def\Ceff#1   {\ensuremath{\mathcal{C}_{#1}^{\mathrm{(eff)}}}\xspace}        % 9  
\def\Cpeff#1  {\ensuremath{\mathcal{C}_{#1}^{'\mathrm{(eff)}}}\xspace}       % 7
\def\Ope#1    {\ensuremath{\mathcal{O}_{#1}}\xspace}                       % 2
\def\Opep#1   {\ensuremath{\mathcal{O}_{#1}^{'}}\xspace}                    % 7

\newcommand{\tev}{\ensuremath{\mathrm{\,Te\kern -0.1em V}}\xspace}
\newcommand{\gev}{\ensuremath{\mathrm{\,Ge\kern -0.1em V}}\xspace}
\newcommand{\mev}{\ensuremath{\mathrm{\,Me\kern -0.1em V}}\xspace}
\newcommand{\kev}{\ensuremath{\mathrm{\,ke\kern -0.1em V}}\xspace}
\newcommand{\ev}{\ensuremath{\mathrm{\,e\kern -0.1em V}}\xspace}
\newcommand{\gevc}{\ensuremath{{\mathrm{\,Ge\kern -0.1em V\!/}c}}\xspace}
\newcommand{\mevc}{\ensuremath{{\mathrm{\,Me\kern -0.1em V\!/}c}}\xspace}
\newcommand{\gevcc}{\ensuremath{{\mathrm{\,Ge\kern -0.1em V\!/}c^2}}\xspace}
\newcommand{\gevgevcccc}{\ensuremath{{\mathrm{\,Ge\kern -0.1em V^2\!/}c^4}}\xspace}
\newcommand{\mevcc}{\ensuremath{{\mathrm{\,Me\kern -0.1em V\!/}c^2}}\xspace}

\def\invfb   {\ensuremath{\mbox{\,fb}^{-1}}\xspace}

\def\invab   {\ensuremath{\mbox{\,ab}^{-1}}\xspace}

\def\gsim{{~\raise.15em\hbox{$>$}\kern-.85em
          \lower.35em\hbox{$\sim$}~}\xspace}
\def\lsim{{~\raise.15em\hbox{$<$}\kern-.85em
          \lower.35em\hbox{$\sim$}~}\xspace}

\newcommand{\Real}{\ensuremath{\mathcal{R}e}\xspace}
\newcommand{\Imag}{\ensuremath{\mathcal{I}m}\xspace}

\def\sqs   {\ensuremath{\protect\sqrt{s}}\xspace}

\def\degrees{\ensuremath{^{\circ}}\xspace}

\newcommand{\lum} {\ensuremath{\mathcal{L}}\xspace}
\def\tell1  {TELL1\xspace}
\def\ukl1   {UKL1\xspace}

\newcommand{\belleTwo}{Belle~II\xspace}
\newcommand{\belleOneTwo}{Belle~(II)\xspace}
\newcommand{\Rd}{\ensuremath{R(D)}\xspace}
\newcommand{\Rdst}{\ensuremath{R(\Dstar)}\xspace}
\newcommand{\Rdordst}{\ensuremath{R(D^{(*)})}\xspace}
\newcommand{\Rk}{\ensuremath{R(\kaon)}\xspace}
\newcommand{\Rkst}{\ensuremath{R(\kaon^*)}\xspace}

\newcommand{\Cnine}{\ensuremath{   {C^{\rm NP}_9}^{\mu\mu}  }   \xspace}
\newcommand{\Cten}{\ensuremath{ {C^{\rm NP}_{10}}^{\mu\mu}   }   \xspace}
\newcommand{\Cnineee}{\ensuremath{ {C^{\rm NP}_9}^{ee}    }   \xspace}
\newcommand{\Cninep}{\ensuremath{ {C^{\prime}_{9}}^{\mu\mu}  }   \xspace}
\newcommand{\Ctenp}{\ensuremath{  {C^{\prime}_{10}}^{\mu\mu}   }   \xspace}
\newcommand{\ReCsev}{\ensuremath{ \Real \left ( C^{\rm NP }_7\right )}\xspace}
\newcommand{\ImCsev}{\ensuremath{ \Imag \left ( C^{\rm NP }_7 \right )}\xspace}
\newcommand{\ReCsevp}{\ensuremath{ \Real \left ( C^{\prime \, \rm NP}_7\right )}\xspace}
\newcommand{\ImCsevp}{\ensuremath{ \Imag \left ( C^{\prime \, \rm NP}_7 \right )}\xspace}

\newcommand{\ADeltaG}{{\ensuremath{{\mathcal{A}}^\DG}}\xspace}
\newcommand{\SKstgamma}{{\ensuremath{S_{\Kstar\g}}}\xspace}
\newcommand{\ATtwo}{{\ensuremath{A^{(2)}_T}}\xspace}
\newcommand{\ATIm}{{\ensuremath{A^{\rm Im}_T}}\xspace}

\def \btosll     {\decay{\bquark}{\squark\ellell}}
\def \btosgamma     {\decay{\bquark}{\squark\g}}
\def \btocell    {\decay{\bquark}{\cquark \ell \neulb}}
\def \btouell    {\decay{\bquark}{\uquark \ell \neulb}}
\def \btoctau    {\decay{\bquark}{\cquark \tauon \neut}}

\def \Lbpmunu    {\decay{\Lb}{\proton \mun \neu}}
\def \LbLcmunu   {\decay{\Lb}{\Lc \mun \neu}}
\def \LcpKpi     {\decay{\Lc}{\proton\Kp\pim}}

\def \Bsmumu     {\decay{\Bs}{\mup\mun}}
\def \Bsll     {\decay{\Bs}{\ell^+ \ell^-}}
\def \Bstautau     {\decay{\Bs}{\taup\taum}}
\def \Bsee     {\decay{\Bs}{\ep\en}}
\def \Bsemu     {\decay{\Bs}{\epm\mump}}
\def \Bdmumu     {\decay{\Bd}{\mup\mun}}
\def \Bdtautau     {\decay{\Bd}{\taup\taum}}
\def \Bdee     {\decay{\Bd}{\ep\en}}
\def \Bdemu     {\decay{\Bd}{\epm\mump}}
\def \tautothreemu {\decay{\tauon^{-}}{\mu^{+}\mu^{-}\mu^{-}}}
\def \tautomugamma {\decay{\tauon^{-}}{\mu^{-}\gamma}}
\def \BtoKpnunu	{\decay{\Bp}{\Kp\neu\neub}}
\def \BtoKstpnunu	{\decay{\Bp}{\Kstarp\neu\neub}}
\def \BtoKstznunu	{\decay{\Bz}{\Kstarz\neu\neub}}

\def \BXlnu      {\decay{\B}{X\ell\neu}}
\def \BXll       {\decay{\B}{X\ellell}}

\def \BDmunu     {\decay{\Bz}{\Dstarp \mun \neumb}}

\def \BPilnu     {\decay{\B}{\pi \ell \neu}}
\def \BDlnu      {\decay{\B}{\D \ell \neu}}
\def \Bdpilnu    {\decay{\Bd}{\pim\ellp\neu}}
\def \Bmtaunu    {\decay{\Bm}{\taum\neub}}

\def \Bstophigamma     {\decay{\Bs}{\phi\g}}
\def \BtoKorKstmumu      {\decay{\B}{\kaon^{(*)}\mup\mun}}
\def \BtoKorKstee      {\decay{\B}{\kaon^{(*)}\ep\en}}
\def \BtoKstmumu      {\decay{\Bz}{\Kstarz\mup\mun}}
\def \BtoKstee      {\decay{\Bz}{\Kstarz\ep\en}}
\def \BtoKmumu      {\decay{\Bp}{\Kp\mup\mun}}

\def \Bstophimumu      {\decay{\Bs}{\phi\mup\mun}}

\def \BtoKstpgamma	{\decay{\Bp}{\Kstarp\g}}
\def \BtoKstzgamma	{\decay{\Bz}{\Kstarz\g}}
\def \BtoXsgamma	{\decay{\B}{X_s\g}}
\def \BXsmumu      {\decay{\B}{X_s\mup\mun}}
\def \BXsll      {\decay{\B}{X_s\ellell}}

\begin{document}

\title{Future prospects for exploring present day anomalies in flavour physics measurements with Belle~II and LHCb}

\author[dor]{J.~Albrecht}
\author[kit,bon]{F.~U.~Bernlochner}
\author[cam]{M.~Kenzie\corref{cor1}}
\ead{matthew.kenzie@cern.ch}
\author[dor]{S.~Reichert\corref{cor1}}
\ead{stefanie.reichert@cern.ch}
\author[ecu]{D.~M.~Straub}
\author[cam]{A.~Tully}

\cortext[cor1]{Corresponding author}

\address[dor]{Fakult\"{a}t Physik, Technische Universit\"{a}t Dortmund, Dortmund, Germany}
\address[kit]{Institut f\"ur Experimentelle Kernphysik, KIT, Karlsruhe,~Germany}
\address[bon]{Physikalisches Institut der Rheinischen Friedrich-Wilhelms-Universit\"{a}t, Bonn, Germany}
\address[cam]{Cavendish Laboratory, University of Cambridge, Cambridge, UK}
\address[ecu]{Excellence Cluster Universe, TUM, Garching, Germany}

\begin{abstract}
A range of flavour physics observables show tensions with their corresponding Standard Model expectations:
measurements of leptonic flavour-changing neutral current processes and ratios of semi-leptonic branching fractions
involving different generations of leptons show deviations of the order of four standard deviations. If confirmed, either
would be an intriguing sign of new physics.
In this manuscript, we
analyse the current experimental situation of such processes and
for the first time estimate the combined impact of the future datasets of the \belleTwo and \lhcb experiments on
the present tensions with the Standard Model expectations by performing scans
of the new physics contribution
to the Wilson coefficients.
In addition, the present day and future
sensitivity of tree-level CKM parameters, which offer orthogonal tests of the Standard Model, are explored.
Three benchmark points in time are chosen for a direct comparison of the estimated sensitivity between the experiments.
A high complementarity between the future sensitivity achieved by the \belleTwo and \lhcb experiments is observed due
to their relative strengths and weaknesses. We estimate that all of the anomalies considered here will be either
confirmed or ruled out by both experiments independently with very
high significance by the end of data-taking at \belleTwo and the
\lhcb upgrade.
%\keywords{}
\end{abstract}


\maketitle

\renewcommand{\arraystretch}{1.25}

\section{Introduction}
\label{sec:intro}

The study of precision observables at the \B factory experiments \babar~\cite{BaBarExperiment} and \belle~\cite{BelleExperiment}, as well as at \lhcb~\cite{LHCbExperiment}
have gained renewed
attention recently: several measurements challenge the lepton flavour universality\\(LFU) assumed in the
Standard Model (SM) and show apparent deviations in the electroweak couplings between the three generations of leptons,
see e.g.~Refs.~\cite{Ciezarek:2017yzh,Archilli:2017} for two recent reviews. In this manuscript, the present day and
future sensitivities of these signatures at the upcoming flavour factory \belleTwo~\cite{Belle2TDR} and the planned running periods of \lhcb and its upgrade~\cite{LHCbFrameworkTDR}
are discussed. For the first time the future impact on the combination of both experiments is investigated in this document.
In addition, studies are performed to assess the future improvements in precision measurements of observables relating to the
Cabibbo-Kobayashi-Maskawa (CKM)
quark mixing matrix~\cite{Cabibbo:1963yz,Kobayashi:1973fv}, a stringent test for over-constraining the SM under the CKM unitarity assumption.
Specifically, the following observables have been studied:

\begin{itemize}
\item[1.]
The CKM parameters \Vub, \Vcb, and $\gamma$: These observables are often
considered as benchmarks for the SM because they can be determined exclusively by tree-level processes. Any disagreement
between tree and loop determinations of the CKM parameters would be clear evidence of new physics (NP) participating in
either the tree or loop diagrams (or both). Furthermore, measurements of \Vub and \Vcb using
different methods show a long-standing tension, diminishing the full potential of this test.
\item[2.] The ratio of tree-level semi-tauonic decays compared to the $\ell = e, \mu$ light lepton final states: This
ratio has been measured in \btoctau decays with \D an \Dstar final
states~\cite{Lees:2012xj,Lees:2013uzd,Huschle:2015rga,Sato:2016svk,Hirose:2016wfn,Aaij:2015yra,Aaij:2017uff} reconstructed using either the leptonic or hadronic \tauon
decay final states. LFU violation is expected in the presence of new phenomena contributing to these processes such as leptoquarks
or new charged currents. The latest experimental measurements of these ratios show a deviation from the SM expectation of about 4$\sigma$,
where $\sigma$ is the standard deviation.
\item[3.] Flavour-changing neutral current decays: Decays involving $\decay{\bquark}{\squark}$ transitions are
loop- and CKM suppressed in the SM and thus exceptionally sensitive to NP. A large number of measurements has been
performed, first by the \B factories, then at the Large Hadron Collider (LHC), in particular by the \lhcb experiment. While no conclusive evidence
for physics beyond the SM has emerged, numerous deviations from SM predictions have been observed.
Discrepancies in tests of lepton flavour universality at the level of $2$~--~$3\sigma$ have been measured in the ratios between
\BtoKorKstmumu and \BtoKorKstee branching fractions~\cite{Aaij:2014ora, Aaij:2017vbb} and further tensions between the SM
predictions and measurements of \btosll transitions~\cite{Aaij:2014pli, Aaij:2016cbx, Aaij:2015oid, Wehle:2016yoi,
Aaij:2015esa, Aaij:2015xza} point towards a deficit in branching
fractions involving decays with muons in the final state. %Further anomalies are present in \Bsdtoll~\cite{Aaij:2017vad, Aaboud:2016ire}.
\end{itemize}

For the projections of the future \belleTwo and \lhcb sensitivity,
published measurements are extrapolated to the relevant anticipated
event yields, incorporating the expected integrated luminosity, production rates
and reconstruction efficiency. Various sources of experimental and theoretical systematic uncertainty are
also estimated and included where appropriate.
In the case of \lhcb, the $b\bar{b}$-production cross-sections
are assumed to scale linearly with centre-of-mass energy, $\sqrt{s}$, as indicated
by the existing measurements \cite{Aaij:2013noa,Aaij:2016avz}.
The trigger efficiencies of the LHCb upgrade are assumed to improve
due to the implementation of a full software trigger according to
Ref.~\cite{CERN-LHCC-2014-016}. Although most presented extrapolations are novel, a small subset of future extrapolations for LHCb are taken
from Ref.~\cite{LHCbFrameworkTDR}. In the case of \belleTwo many extrapolations
are also taken from Ref.~\cite{BelleIIBook}, which provides a full overview
of the \belleTwo physics program and potential.

\begin{figure*}[tbhb!]
     \centering
           \includegraphics[width=0.98\textwidth]{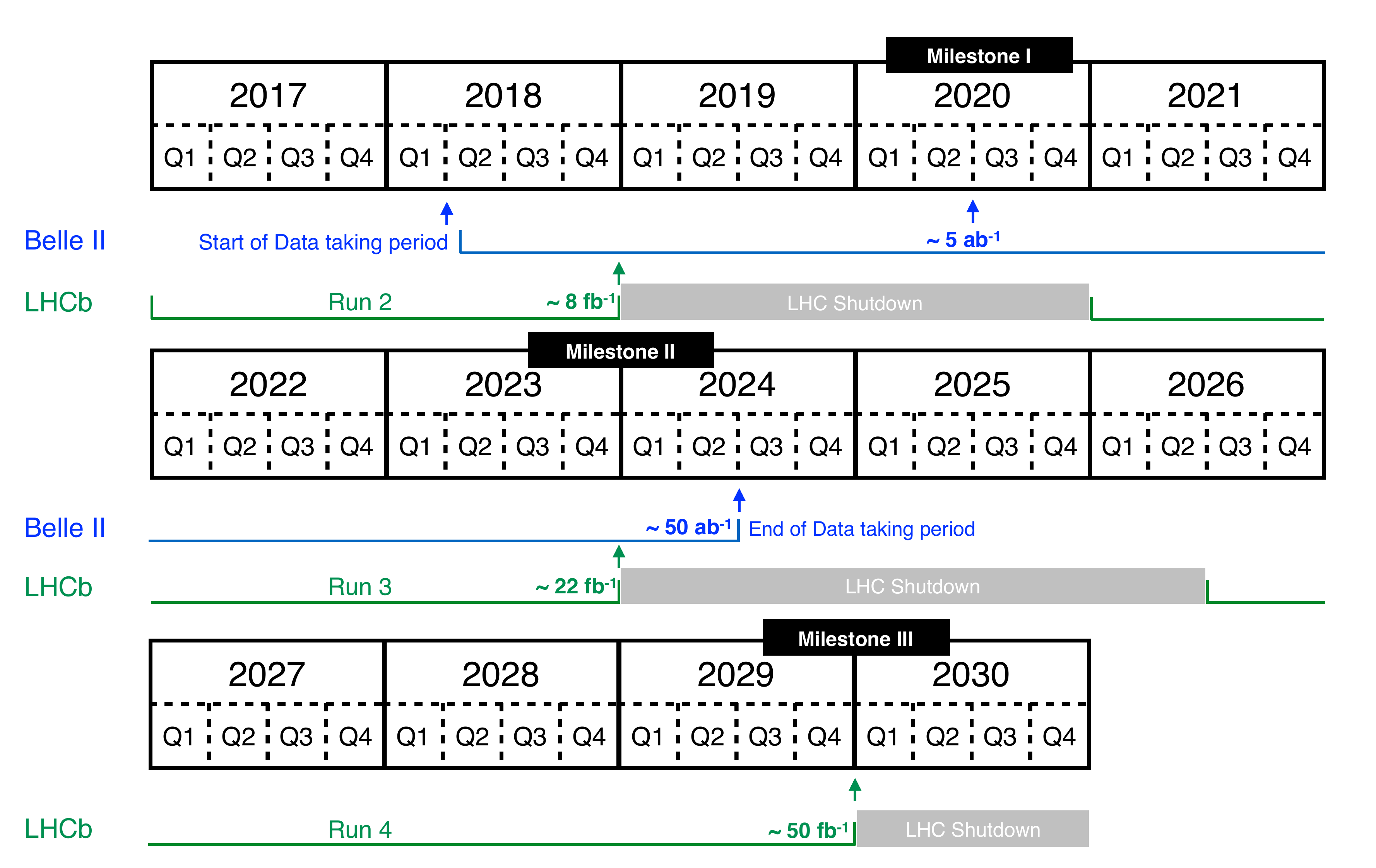}
      \caption{
      An overview of the expected \belleTwo and \lhcb timelines along with their estimated integrated luminosities at each milestone.
      The scenarios compared in this manuscript are shown in bold.
      For more details of the expected luminosities and number of
      produced $\bquark\bquarkbar$-pairs at each milestone see Table~\ref{table:lumis}.
      The LHCb phase~1 upgrade~\cite{CERN-LHCC-2014-016} is currently scheduled for the
      duration of the LHC shutdown between 2019~--~2020. The LHCb experiment has recently expressed its interest to continue running past the phase~1 upgrade
      until the end of the funded LHC Run in 2035~\cite{Aaij:2244311}.
       }
     \label{fig:timeline}
 \end{figure*}

An overview of the expected \belleTwo and \lhcb data taking periods is
given in Figure~\ref{fig:timeline}. Using these periods, three
`milestone' points are chosen to provide estimated sensitivities in
the years 2020, 2024 and 2030. They are summarised in
Table~\ref{table:lumis} and defined as follows:
The Belle experiment has recorded 0.7\invab at the $\Upsilon(4S)$ centre-of-mass energy throughout its running in the
years 1999~--~2010 and the \lhcb experiment has recorded 3\invfb in Run~1 (2010~--~2012) of the LHC.  The \belleTwo experiment
plans to record collisions at the $\Upsilon(4S)$ centre-of-mass energy in 2018 and will have accumulated a dataset of $\sim5\invab$ by 2020,
which corresponds to an
approximately four-fold increase in the size of the dataset from the existing \B factories.
A comparable point in time for the \lhcb experiment is the scheduled end of Run~2 of the LHC (2015~--~2018) by
which time approximately 8\invfb of $pp$ collision data will have been collected.
These two datasets define the first `milestone'.
By mid 2024 \belleTwo will have accumulated its full envisioned $\Upsilon(4S)$ dataset of about 50\invab. In this time
the \lhcb phase~1 upgrade plans to have recorded
a dataset of 22\invfb using collisions from Run~3 (2021~--~2023) of the LHC. This defines the second `milestone'.
As of now, no concrete plans exist for a possible \belleTwo
upgrade, thus no assumptions beyond 2024 are made for \belleTwo.
The \lhcb phase~1 upgrade plans to collect a total dataset of 50\invfb
by 2029 using collisions from Run~4 (2026~--~2029) of the LHC, marking the third `milestone' scenario. The timelines of the
future data taking of the \belleTwo and \lhcb experiments are taken
from Ref.~\cite{BelleIIBook} and from the LHC
roadmap~\cite{MTP20162020V1}.

The \lhcb experiment has recently expressed its interest to
continue running after the phase~1 upgrade~\cite{Aaij:2244311}.
This phase~2 upgrade is suggested to run until the end of the funded
LHC Run, 2035, and will collect a total dataset of 300\invfb.
The phase~2 upgrade of \lhcb is not further discussed in this document. The \belleTwo experiment plans also to record
datasets beyond the $\Upsilon(4S)$ centre-of-mass energy, which will allow for studies of higher mass states, for example the $B_s$-meson. As these plans are not
detailed at this point in time, we will not further discuss them in this document.

\begin{table*}[tbht!]
\caption{
    The luminosity scenarios considered along with the estimated number of $\bquark\bquarkbar$-pairs produced inside the
    acceptance of the experiments are given. The \lhcb cross sections are taken from Ref.~\cite{Aaij:2013noa} assuming a
    linear increase in $\bquark\bquarkbar$-production cross section with LHC beam energy. For \belleTwo only
    $e^+ e^- \to \Upsilon(4S) \to B \bar B$ data sets are estimated. \label{table:lumis}
}
\begin{center}\begin{tabular}{lcccccc}
\hline
  &                      &   & ~~~~~`Milestone~I'~~~~~&~~~~~`Milestone~II'~~~~~&~~~~~`Milestone~III'~~~~~\\
year         &                       & 2012 & 2020 & 2024& 2030 \\
%LHC run   &                        & ~~~~~1~~~~~ & ~~~~~2~~~~~ & ~~~~~3~~~~~ & ~~~~~4~~~~~ \\
\hline
\lhcb& \lum [\invfb]&      3 &8&22&50\\
         &  n($\bquark\bquarkbar$) &     $0.3 \times 10^{12}$ & $1.1 \times10^{12}$&$37 \times10^{12}$&$87 \times10^{12}$\\
         & $\sqs$ & $7/8\tev$ & $13\tev$ & $14\tev$ & $14\tev$\\
\hline
\belleOneTwo & \lum [\invab]&    0.7 &5&50& - \\
          &  $n( B \bar B)$&      $0.1 \times10^{10}$ &$0.54 \times 10^{10}$&$5.4 \times 10^{10}$  & -\\
          & $\sqs$ & $10.58 \gev$ &  $10.58 \gev$  & $10.58 \gev$  & -\\

\hline
\end{tabular}\end{center}
\end{table*}

The remainder of this manuscript is structured as follows: In Section~\ref{sec:CKM}, tests of the CKM structure of the SM are
detailed. Section~\ref{sec:RdRdst} discusses the test of LFU in tree level semi-leptonic decays. Section~\ref{sec:LFU} focuses on
flavour-changing neutral current decays based on quark-level $\decay{\bquark}{\squark}$ transitions, including loop-level
tests of LFU and a study of the impact of the future experiments on the knowledge of the Wilson
coefficients. The manuscript concludes with Section~\ref{sec:conclusion} which summarises the main findings.

\section{Measurements of tree-level CKM parameters}
\label{sec:CKM}

The flavour structure, and consequent phenomena of \CP violation, in the quark sector of the SM are completely described by
the CKM quark mixing matrix~\cite{Cabibbo:1963yz,Kobayashi:1973fv}. A common representation
of the CKM constraints is portrayed in the complex plane as the so-called unitarity triangle, which has a single
unknown apex $(\rhobar, \etabar)$. This point can be uniquely determined using the length of the side opposite the
well-measured angle $\beta$, which is proportional to the ratio $\Vub/\Vcb$, and the relatively poorly-known angle
$\gamma$. Both of these can be determined from tree-level decays.
Under the SM hypothesis, the apex determined using $\gamma$ and $\Vub/\Vcb$ should be the same as that
determined from $\beta$ and $\Delta m_{d}/\Delta m_{s}$, which are determined from loop decays.
Given the latter are considerably better measured than the
former, precision determinations of $\gamma$, \Vub and \Vcb are important tests of the CKM structure of the SM.
It has been shown that the current
experimental constraints on the Wilson coefficients governing decays of the form
$\decay{\bquark}{\uquark_{1}\uquarkbar_{2}\dquark_{1}}$, where $\uquark_{1,2}$ are up-type quarks and $\dquark_{1}$
is a down-type quark, can still easily allow for tree-level new physics effects of order 10\%~\cite{Brod:2014bfa}.
Effects of this size can cause shifts in the tree-level determination of \g of up to $4^\circ$. Thus, comparison between
the point in $(\rhobar, \etabar)$ space determined using $\g$ and $\Vub/\Vcb$ with that found using $\sin(2\beta)$
and $\dmd/\dms$ is a cornerstone of the flavour physics program at both \lhcb and \belleTwo, where any discrepancies
will be of huge importance.

Sensitivity to \Vub and \Vcb arises from the semileptonic transitions \btouell and \btocell respectively. This can be
achieved with two different analysis techniques; using either inclusive or exclusive final states. Exclusive
measurements use specific decay modes which proceed via a $\decay{\bquark}{\uquark}$ or $\decay{\bquark}{\cquark}$
transition, for example \BPilnu~\cite{Sibidanov:2013rkk} or \BDlnu~\cite{Glattauer:2015teq}, to determine \Vub and \Vcb
respectively.\footnote{Charge conjugation is implied throughout.} These require experimental
extraction of the differential decay rate along with theoretical input parameterising the form factor. Inclusive
measurements use the sum of all possible decays of the type \btouell and \btocell, for \Vub and \Vcb respectively. These
are experimentally more challenging due to considerable background contamination which can only be removed by
restriction to a particular region of the available phase space.
Extraction of \Vub and \Vcb in the hadronic environment of the LHC is extremely challenging, if
not impossible, using the channels described above. Instead, \lhcb has pioneered a new approach in which the ratio
$\Vub/\Vcb$ is extracted using baryonic decays of \Lbpmunu relative to \LbLcmunu~\cite{Aaij:2015bfa}. This is equivalent
to an exclusive determination and requires input from lattice QCD of the relevant form factors.

\begin{figure}
    \centering
    \subfigure[Progression of inclusive (square points) and exclusive (circular points) measurements of \Vub (red) and \Vcb (blue)]{
      \includegraphics[width=0.485\textwidth]{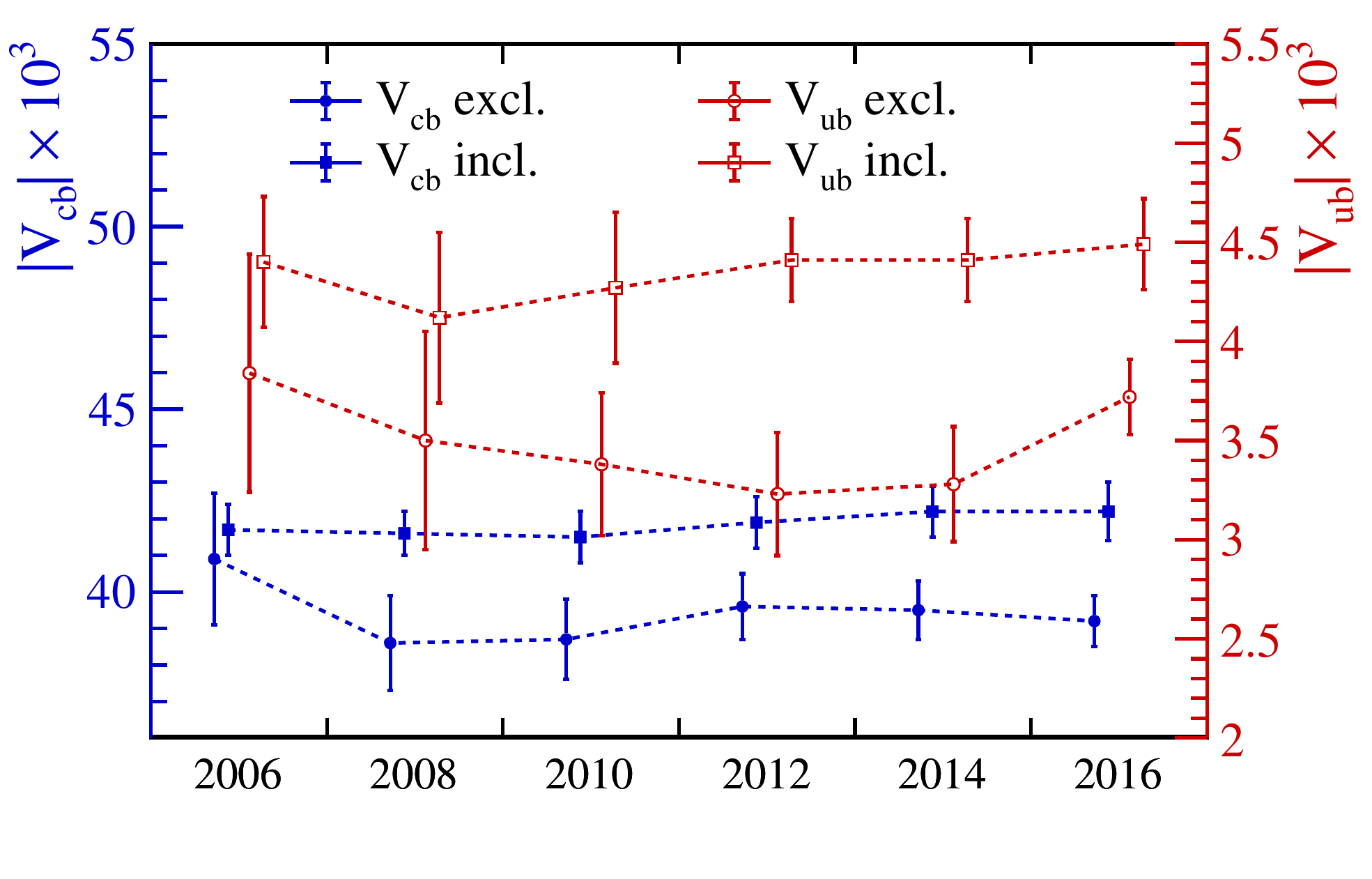}
      \label{fig:VubVcbValues}
    }
    \hspace{0.002\textwidth}
    \subfigure[Compatibility, in standard deviations ($\sigma$), between inclusive and exclusive measurements of \Vub (red) and \Vcb (blue)]{
      \includegraphics[width=0.455\textwidth]{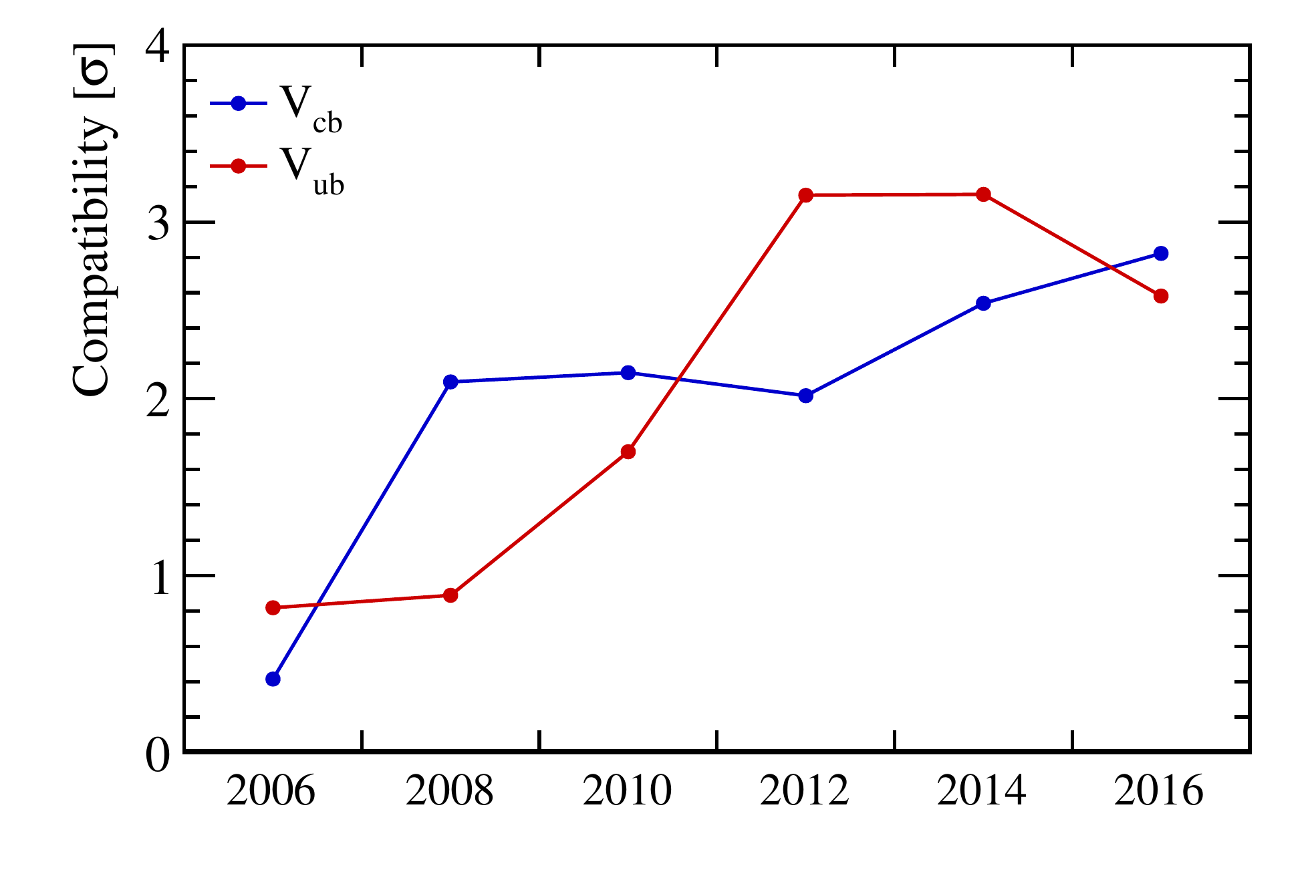}
      \label{fig:VubVcbSigma}
    }
    \caption{Historical progression of inclusive and exclusive measurements of \Vub and \Vcb from Ref.~\cite{PDG2016}.}
    \label{fig:VubVcb_progression}
\end{figure}

Historically, there has been something of a puzzle surrounding these measurements
because of a long
standing discrepancy between the
inclusive and exclusive approaches. This is illustrated in Fig.~\ref{fig:VubVcb_progression} (numbers sourced
from Ref.~\cite{PDG2016}). Resolving and understanding these discrepancies, whether physically motivated by new physics
or due to theoretical or experimental oversights, is an important goal for upcoming flavour physics experiments.

\begin{figure}
    \centering
    \includegraphics[width=0.7\textwidth]{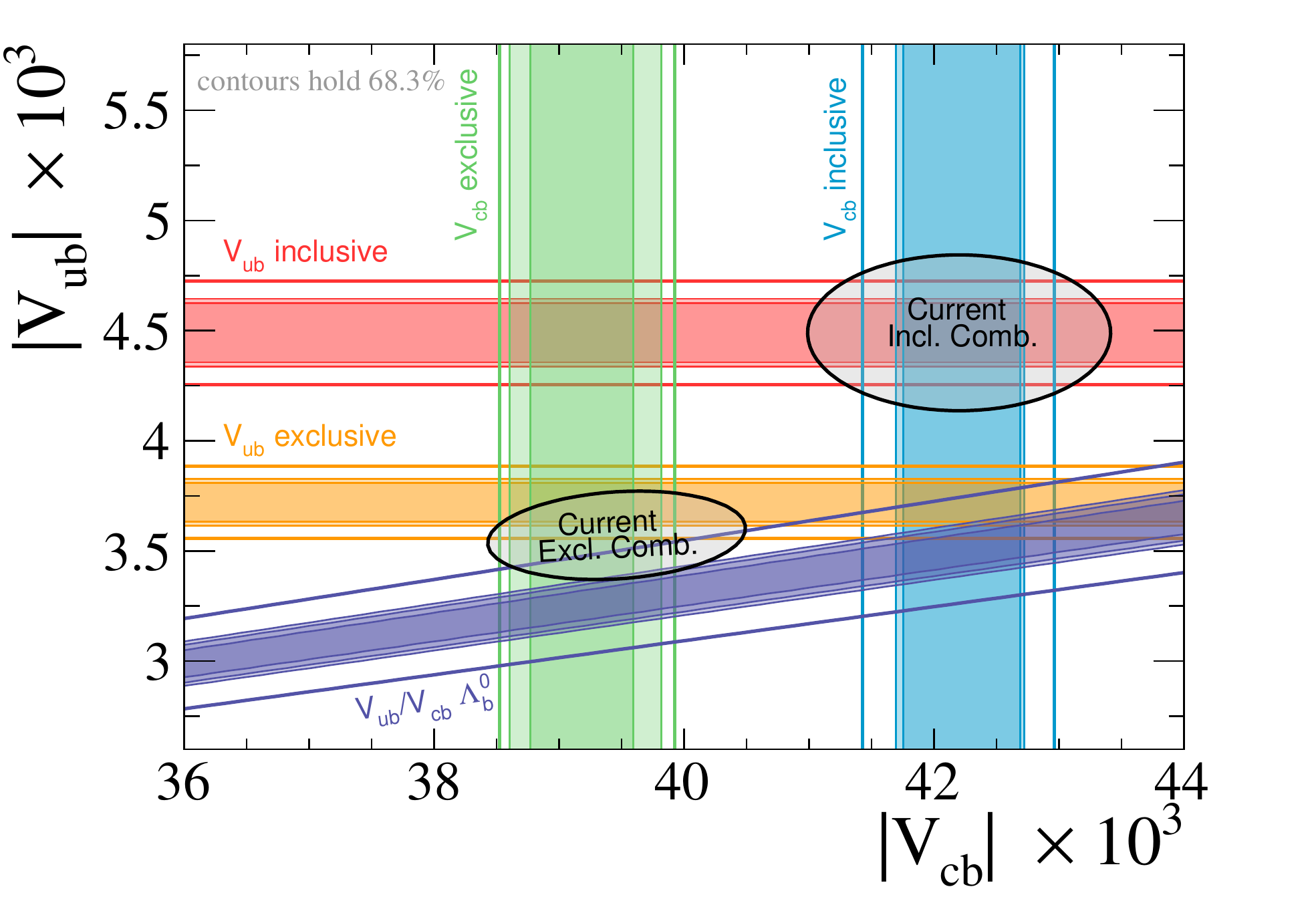}
    \caption{Prospects of the future sensitivity for various inclusive and exclusive measurements of \Vub and \Vcb. Each colour represents a different
      measurement: \Vub inclusive (red), \Vub exclusive (orange), \Vcb inclusive (cyan), \Vcb exclusive (green) and the $\Vub/\Vcb$ \Lb analysis (blue).
      The current world averages,
      from Ref.~\cite{PDG2016}, are shown by the solid lines with no fill. The future projections at milestones I, II and III are given by the
      filled contours and are progressively overlaid. The current inclusive and exclusive
      combinations are shown as the gray filled areas.}
    \label{fig:VubVcb}
\end{figure}

\begin{figure}
  \centering
  \includegraphics[width=0.7\textwidth]{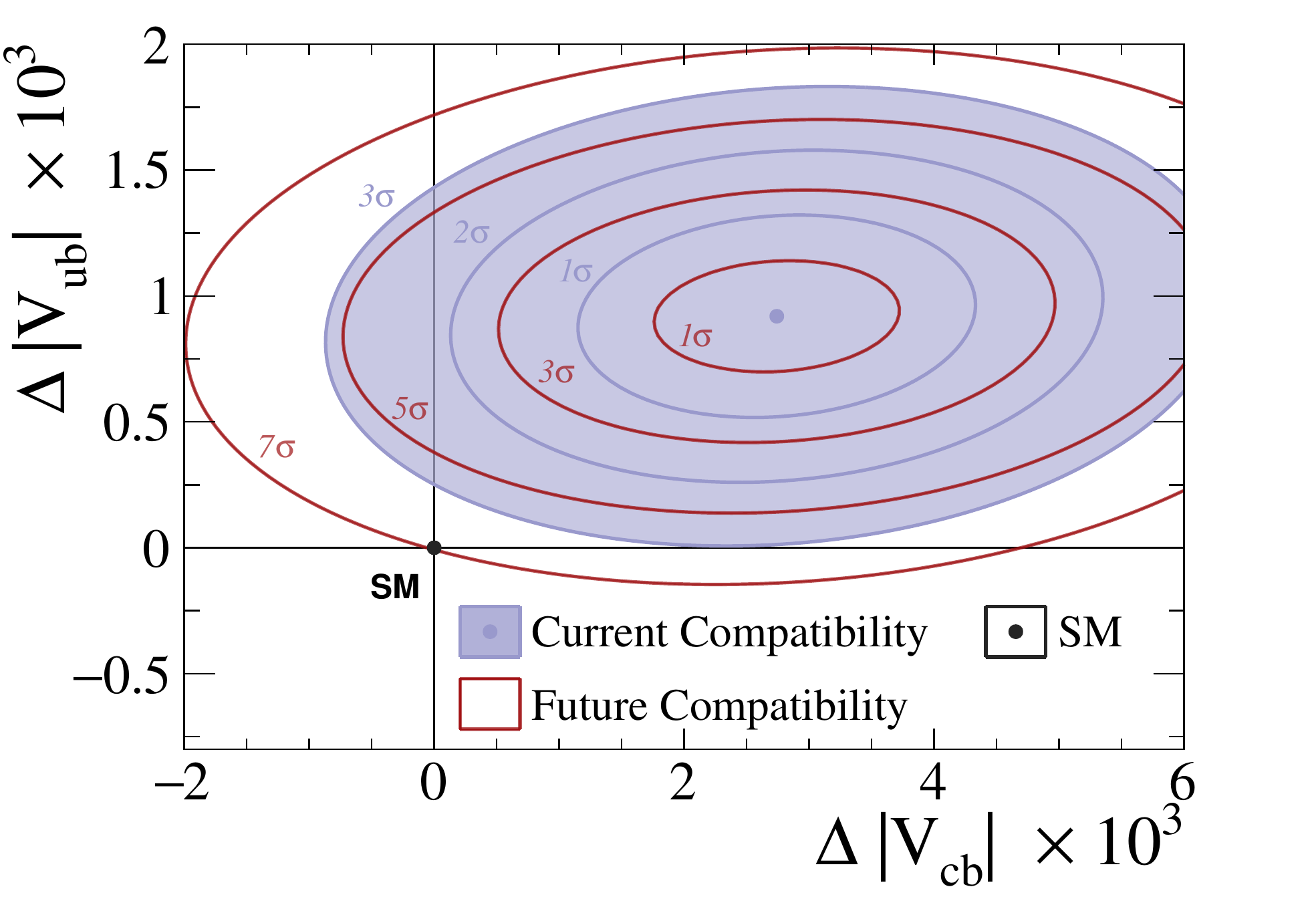}
  \caption{The current (blue) and future (red) compatibility, at milestone III, of the difference between inclusive and exclusive measurements of \Vub and \Vcb.}
  \label{fig:VubVcbComp}
\end{figure}

In this
section the impact of upcoming data from the \belleTwo~\cite{Belle2TDR} and \lhcb
experiments~\cite{LHCbFrameworkTDR} is estimated, where the current world averages from Ref.~\cite{PDG2016} are used for the central values.
Estimates for the future projections of the uncertainties for \belleTwo come from Ref.~\cite{BelleIIBook}, whilst
for the $\Vub/\Vcb$ ratio measurement from \lhcb the experimental uncertainty is reduced as detailed in the introduction by the square-root of the
expected yield increase which arises from the increased luminosity, collision energy and trigger improvement.
It has been assumed that knowledge of the \LcpKpi branching fraction, which will become the dominant experimental systematic uncertainty when
extracting the absolute value of \Vub from the ratio $\Vub/\Vcb$ using \Lb decays, improves inline with the increased
dataset expected at \belleTwo.
Furthermore, we assume modest improvement to the
uncertainty arising from lattice QCD.
The values used are shown in Table~\ref{tab:VubVcbNumbers} and the projection of the future impact shown in
Fig.~\ref{fig:VubVcb}, using the \texttt{GammaCombo} package~\cite{gammacombo}.
Figure~\ref{fig:VubVcb} demonstrates that the largest improvement for inclusive measurements comes from the first
milestone of \belleTwo, with relatively little further impact from the second milestone. Conversely, for exclusive
measurements at \belleTwo the big improvement, especially for \Vcb, arises from the large increase in the sample size of
\belleTwo at the second milestone. For the \lhcb baryonic measurement the biggest improvement is due to the increase in
luminosity for the first milestone but then becomes limited by the systematic uncertainties from lattice QCD.
Future improvements to these uncertainties will be vital in order to extract the best possibile sensitivity
from the baryonic measurement of the $\Vub/\Vcb$ ratio.
Figure~\ref{fig:VubVcbComp} shows the
compatibility of the current measurements, and future projections, with the SM by examining the difference between inclusive and exclusive
measurements.
This demonstrates that, if the current central values stay the same, the
discrepancy will be well beyond $5\sigma$.
An additional figure in~\ref{sec:appendix_figs}, Fig.~\ref{fig:VubVcbB}, shows the compatibility, at $1\sigma$, between
the current world averages for inclusive and exclusive determinations of \Vub and \Vcb, alongside the projected
uncertainties at $1\sigma$, $3\sigma$ and $5\sigma$ when the \belleTwo and
\lhcb experiments have finished running, assuming the central values stay the same.
\begin{table*}
  \begin{center}
  \caption{Values used for the projections of future \Vub and \Vcb measurements}
  \begin{tabular}{l c c c c c c c}
    \hline
    Measurement     &  Current World                     & Current            & \multicolumn{5}{c}{Projected Uncertainty } \\
                    &  ~~~Average ($\times10^{-3}$)~~~   & ~~~Uncertainty~~~  & \multicolumn{2}{c}{\belleTwo}   & \multicolumn{3}{c}{\lhcb}           \\
                    &  \small{(Ref.~\cite{PDG2016})}     &  \small{(Ref.~\cite{PDG2016})} &  5\invab &  50\invab &  8\invfb  &  22\invfb  &  50\invfb \\
    \hline
    \Vub inclusive  &  $4.49\pm0.23$                     &  $5.1\%$           & $3.4\%$  & $3.0\%$   & -         & -          & -       \\
    \Vub exclusive  &  $3.72\pm0.19$                     &  $5.1\%$           & $2.5\%$  & $2.1\%$   & -         & -          & -       \\
    \Vcb inclusive  &  $42.2\pm0.8$                      &  $1.9\%$           & $1.3\%$  & $1.2\%$   & -         & -          & -       \\
    \Vcb exclusive  &  $39.2\pm0.7$                      &  $1.8\%$           & $1.6\%$  & $1.1\%$   & -         & -          & -       \\
    $\Vub/\Vcb$     &  $83.0\pm5.7$                      &  $6.9\%$           & -        & -         &  $3.4\%$  &  $2.9\%$   & $2.1\%$ \\
    \hline
  \end{tabular}
  \label{tab:VubVcbNumbers}
  \end{center}
\end{table*}

\begin{figure}
  \centering
  \includegraphics[width=0.7\textwidth]{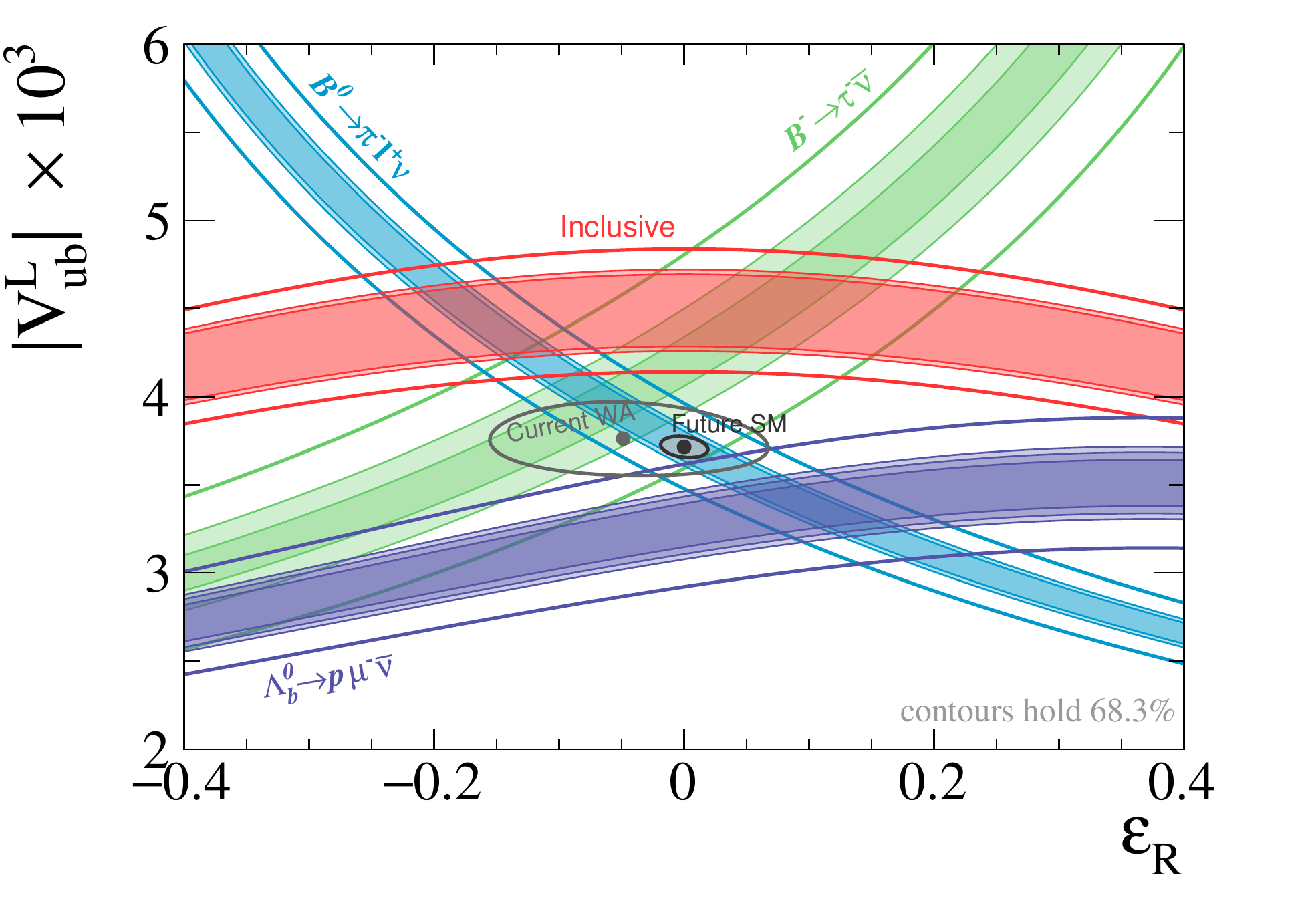}
  \caption{Prospects for new physics measurements related to right-handed currents with the current world averages from Ref.~\cite{PDG2016} (not filled) and the future projections
    at milestones I, II and III (filled) overlaid. The current world average (gray dot and gray line) and the SM point (black dot) with the $1\sigma$ exclusion contour at
  milestone III (black line) are also shown.}
  \label{fig:rh_currents}
\end{figure}

One possible explanation for the discrepancy between inclusive and exclusive measurements of \Vub and \Vcb has been to
introduce a new current which couples only to right-handed fermions \cite{Crivellin:2009sd}. This explanation is probed
by exploring the dependence of \Vub as a left-handed only coupling, $|V_{ub}^{L}|$, on the size of the assumed right-handed
coupling, $\epsilon_R$. Tensions and prospects for these related measurements have been explored using the values shown
in Table~\ref{tab:VubLNumbers}, which are obtained for \belleTwo measurements from~\cite{BelleIIBook}, and using the
same projections from \lhcb for \Lbpmunu as in Table~\ref{tab:VubVcbNumbers}, with an additional source of uncertainty
arising from the \Vcb normalisation term, which is dominated by knowledge of the \LbLcmunu branching fraction. The projections are shown in
Fig.~\ref{fig:rh_currents} in which the central values are taken from the
existing measurements (as listed in Table~\ref{tab:VubLNumbers}). This shows
that the most dramatic improvement comes at the first milestone for the
inclusive, $\BPilnu$ and $\Lbpmunu$ measurements, whilst the $\Bmtaunu$ improvement continues towards milestone III. \ref{sec:appendix_figs} has an additional figure, Fig.~\ref{fig:rh_currents_central}, in which
the central values have been shifted to the SM expectation from Ref.~\cite{CKMfitter} with $\epsilon_R=0$ and
$\Vub=(3.72\pm0.06)\times 10^{-3}$.

It is worth noting some recent development with respect to exclusive \Vcb : The Belle collaboration made the exclusive $B \to D^{*} \, \ell \, \bar \nu_\ell$ unfolded differential decay rates available for the first time in Ref.~\cite{Abdesselam:2017kjf} and analyses using alternative form factors have found a value much more consistent with that of the inclusive analyses.
%It is worth noting that some recent developments regarding the parametrisation of the form factor for the exclusive
%determination of \Vcb from $B\to D^*\ell\bar\nu_\ell$ decays~\cite{Bigi:2017njr,Grinstein:2017nlq} have found a value
%much more consistent with that of the inclusive analyses.
The measurements presented in this document, taken directly
from Ref.~\cite{PDG2016}, use the Caprini-Lellouch-Neubert (CLN) parametrisation~\cite{Caprini:1997mu}, which hitherto has been the one
applied by most experimental analyses (exceptions are e.g. Refs.~\cite{Abreu:2001ic,Aubert:2004bw}). The present level of experimental accuracy calls for a more careful treatment of the associated theoretical uncertainties \cite{Bernlochner:2017jka}. An
alternative parametrisation from Boyd, Grinstein and Lebed (BGL)~\cite{Boyd:1997kz} results in values closer to the
inclusive determination, depending on the particular parameters used. Clearly, deeper understanding of both the
experimental and theoretical approaches is necessary to resolve these discrepancies~\cite{Bernlochner:2017xyx}.

Precision measurements of \Vub and \Vcb can be combined with measurements of the CKM angle $\g$ to determine a uniquely
tree-level measurement of the CKM parameters $(\rhobar, \etabar)$, under the SM hypothesis. This is a good probe for new
physics when compared to measurements of $\sin(2\beta)$, $\dmd$ and $\dms$ which determine the same point from loop
processes. The direct determination of the CKM angle \g predominantly uses decays of the form
$\Bm\to\Dz\Km$ where the ratio between the favoured $\bquark\to\cquark$ and supressed $\bquark\to\uquark$ transitions goes like
$r e^{i(\delta-\gamma)}$, where $r$ and $\delta$ are unknown hadronic parameters. A comprehensive review on the determination of
$\gamma$ can be found in Refs.~\cite{Amhis:2016xyh,Aaij:2016kjh}. Prospects for improved determinations of the CKM angle \g from both \belleTwo and \lhcb
are considerable. By the end of milestone I (II) \belleTwo expect to determine \g with $6\degrees$ ($1.5\degrees$) precision~\cite{BelleIIBook}.
LHCb expect to determine \g at the level of $4\degrees$~(milestone I), $1.5\degrees$~(milestone II) and $<1\degrees$~(milestone III)~\cite{LHCbFrameworkTDR}.

\begin{figure}
  \centering
  \includegraphics[width=0.7\textwidth]{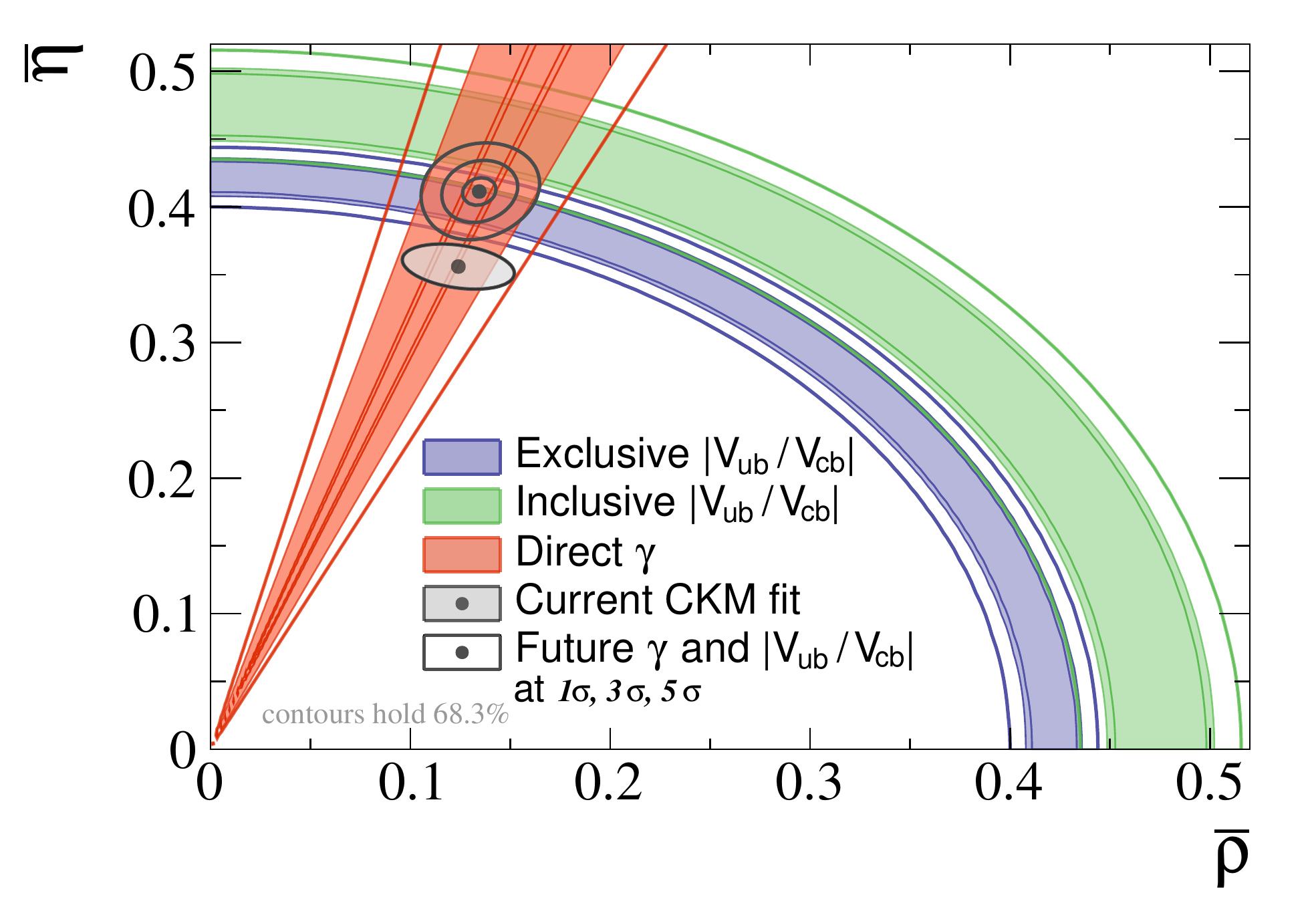}
  \caption{Prospects for CKM fits in $(\rhobar, \etabar)$ space using tree-level processes only with the current world averages from
  Ref.~\cite{PDG2016} (not filled) and the future projections at milestones I, II and III (filled) overlaid. The current CKM fit, using all available constraints, from Ref.~\cite{CKMfitter}
  (gray line with light gray fill) along with the future combination of $\gamma$, \Vub and \Vcb at milestone III (dark gray lines with no fill), where the $1\sigma$, $3\sigma$ and $5\sigma$ exclusion
  contours are shown.}
  \label{fig:RhoEta}
\end{figure}

The projections for exclusive and inclusive determination of \Vub and \Vcb, overlaid with those for direct
determination of CKM angle $\g$, are shown in Fig.~\ref{fig:RhoEta}.
This is overlaid with the current world average using all
contraints on $(\rhobar, \etabar)$ from the CKMfitter collaboration~\cite{CKMfitter}. It it noticeable that already
there is some tension between $\Vub/\Vcb$ measurements and the CKM fit. An additional figure in~\ref{sec:appendix_figs}, Fig.~\ref{fig:RhoEta2}, shows the
same plot with the current experimental constraints on $\sin(2\beta)$ and $\Delta m_{d}/\Delta m_{s}$, from Ref.~\cite{PDG2016}, additionally overlaid.

\begin{table*}
  \begin{center}
  \caption{Values used for the projections of future limits on right-handed currents}
  \begin{tabular}{l c c c c c c c}
    \hline
    Measurement     &  Current World                     & Current            & \multicolumn{5}{c}{Projected Uncertainty } \\
                    &  ~~~Average ($\times10^{-3}$)~~~   & ~~~Uncertainty~~~  & \multicolumn{2}{c}{\belleTwo}   & \multicolumn{3}{c}{\lhcb}           \\
                    &  \small{(Ref.~\cite{PDG2016})}     &  \small{(Ref.~\cite{PDG2016})} &  5\invab &  50\invab &  8\invfb  &  22\invfb  &  50\invfb \\
    \hline
    Inclusive       &  $4.49\pm0.23$                     &  $5.1\%$           & $3.4\%$  & $3.0\%$  & -         & -       & -       \\
    \Bmtaunu        &  $4.2\pm0.4$                       &  $9.5\%$           & $4.7\%$  & $2.2\%$  & -         & -       & -       \\
    \Bdpilnu        &  $3.72\pm0.16$                     &  $4.3\%$           & $2.0\%$  & $1.5\%$  & -         & -       & -       \\
    \Lbpmunu        &  $3.27\pm0.23$                     &  $6.9\%$           & -        & -        &  $3.9\%$  & $3.3\%$ & $2.5\%$ \\
    \hline
  \end{tabular}
  \label{tab:VubLNumbers}
  \end{center}
\end{table*}

\section{Lepton flavour universality in trees}
\label{sec:RdRdst}

A key test of LFU is measuring the ratio of branching fractions of decays that differ only by the lepton content of the
final state. Measurements of this type are represented by the observable $R(X)$, which denotes the ratio of branching fractions of
\BXlnu (or \BXll in the next section) decays, for two choices of $\ell$, where $\ell$ can be $\electron,\muon$ or $\tauon$.

A large class of SM extensions contain new interactions that couple preferentially to the third generation of quarks and
leptons, such as models involving Higgs-like charged scalars or $W'$ bosons. Ratios involving tree-level
$\decay{\bquark}{\cquark\tauon\neu}$ transitions are particularly sensitive to these NP scenarios. Two of these
observables, \Rd and \Rdst, are defined as the ratio of the branching fractions of $\Bd\to\D^{(*)+}\taum\neutb$ to $\Bd\to\D^{(*)+}\ell^{-}\neulb$
with $\ell = e$ or $\mu$. Their Standard
Model predictions are $(0.299\pm 0.003)$ and $(0.257\pm 0.003)$ respectively~\cite{Bernlochner:2017jka} (see also
Refs.~\cite{Fajfer:2012vx,Bigi:2016mdz,Bigi:2017jbd} for other relevant work on this).
Belle and BaBar have made measurements of both \Rd and \Rdst~\cite{Huschle:2015rga, Lees:2012xj, Lees:2013uzd,
Sato:2016svk, Hirose:2016wfn}, while \lhcb has currently only measured \Rdst~\cite{Aaij:2015yra,Aaij:2017uff}.
%\footnote{During the writing of this manuscript, LHCb published a second measurement of \Rdst~\cite{Aaij:2017uff} which is not used for the central value and uncertainty of the current measurement but is
%considered in the future extrapolations.}.
\lhcb also has the potential to measure \Rd, but
has not yet published such a measurement, hence projections for this are not shown.
The HFLAV combination of the \Rdst measurement from \lhcb using muonic \tauon decays with the Belle and BaBar measurements results in a deviation of $3.9\sigma$ from the SM
prediction~\cite{Amhis:2016xyh}.
During the writing of this manuscript, LHCb published a second measurement of \Rdst using hadronic $\tauon$ decays~\cite{Aaij:2017uff}, which was
not included as part of the current world average values in this document.
The addition of this result is expected to shift the central value
of the world average towards the SM predictions slightly, but due to the precision of the measurement, the overall
significance of the deviation stays approximately the same. As this effect is expected to be small, we neglect the addition
of this measurement and proceed with the current HFLAV world average. The hadronic $\tauon$ \Rdst measurement is
considered in the future extrapolations.

Complementary measurements have also been made of the ratios \Rk and \Rkst. These ratios differ from \Rdordst as they do
not occur at tree level in the SM or involve a $\tauon$ lepton and therefore probe NP scenarios that couple to different
generations of fermions in loop processes. Measurements of \Rk and \Rkst by the Belle and BaBar experiments
suffer from small sample sizes~\cite{Wei:2009zv,Aubert:2008ps}, however, the \lhcb measurements of \Rk and \Rkst show
discrepancies with respect to the SM prediction of around $3\sigma$~\cite{Aaij:2014pli,Aaij:2017vbb} and are discussed
in greater detail in Sec.~\ref{sec:LFU}. These measurements, in addition to \Rd and \Rdst, suggest a pattern of
tensions among tests of LFU.

\begin{figure}
  \centering
  \includegraphics[width=0.7\textwidth]{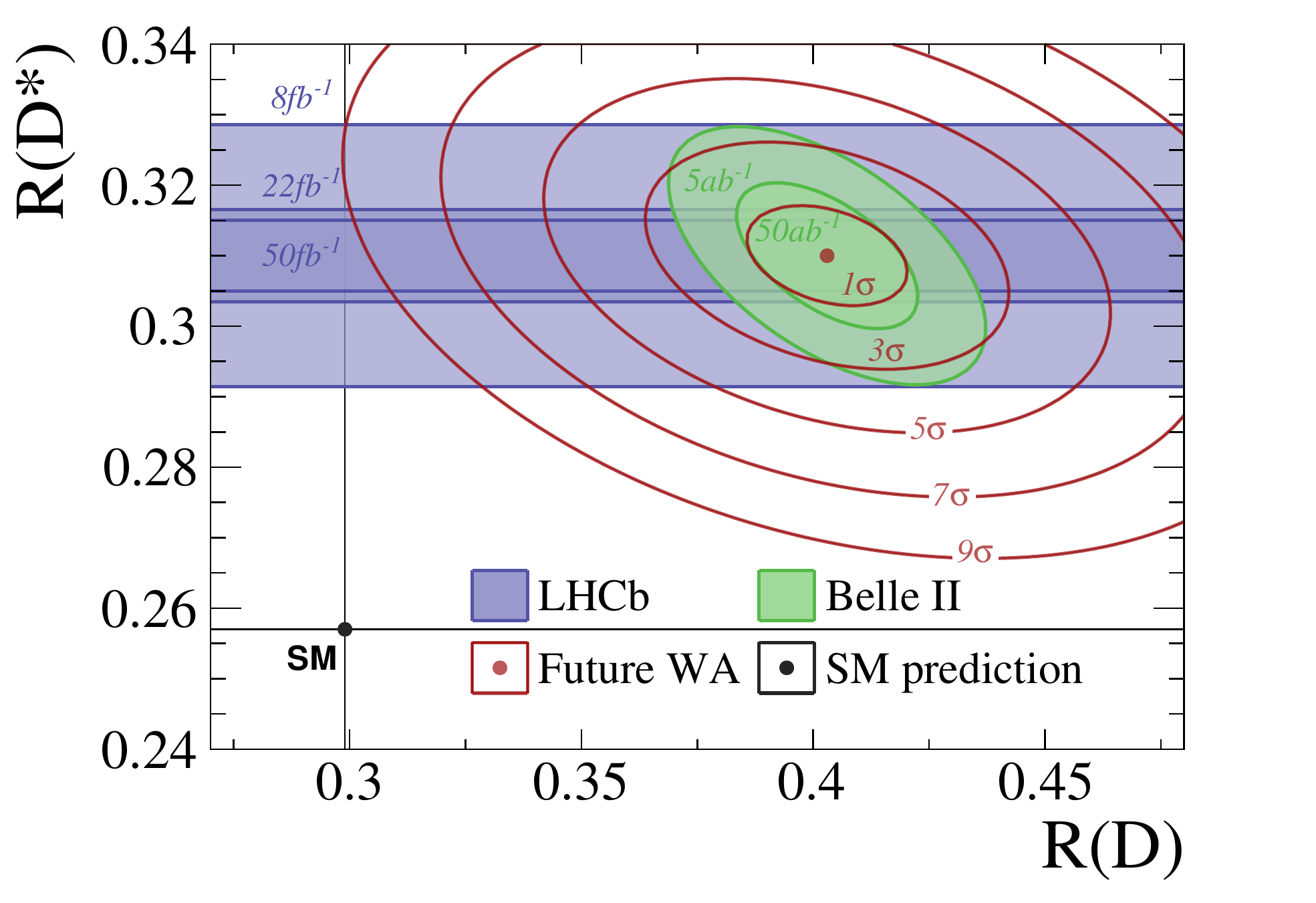}
  \caption{Future prospects for measurements of \Rd and \Rdst. The SM and future
  expected uncertainties at milestone III are combined to predict the significance with which a given point can be excluded if the current central values remain the same (red lines). The
  expected uncertainties from \belleTwo (green) and \lhcb (blue) alone are shown as the shaded bands. The relatively small size of the SM uncertainty compared to the current experimental
  constraints can be seen in Fig.~\ref{fig:RdRdstA}, where the uncertainties are shown separately.}
  \label{fig:RdRdst}
\end{figure}

The large data samples to be collected by the \lhcb and \belleTwo experiments will be sufficient to confirm the
existence of these anomalies, if they are indicative of violation of LFU. In this section, we predict the sensitivity of
\lhcb and \belleTwo to \Rd and \Rdst.
The central values used for the \lhcb and \belleTwo
predictions are taken from the current HFLAV world average~\cite{Amhis:2016xyh}. The \lhcb \Rdst statistical uncertainties are
scaled as detailed previously %from the values measured in the hadronic and muonic channels in Run I according to the expected increase in
integrated luminosity, $B$ production cross section and increase in trigger efficiency~\cite{CERN-LHCC-2014-016}%.
Most of the systematic uncertainties are proportional to the data or control sample size and are scaled in the same way.
However, due to the use of external inputs, there are some irreducible systematic uncertainties. The
external input of the branching fraction of $\decay{\tauon}{\muon\neu\neu}$ to the muonic measurement is not expected to
improve in precision from the measurements made at LEP under ideal conditions for $\tau$ production using
$\decay{\Z}{\tau\tau}$, and hence is kept constant in the future projections at 0.3\%. The hadronic measurement relies
on external input for the branching fractions of $\decay{\Bz}{\Dstarp \pim\pip\pim}$ and \BDmunu, which together
contribute 4.8\% to the systematic uncertainty. The precision of the branching fraction of \decay{\Bz}{\Dstar\muon\neu}
is not expected to change since an independent dataset from the one used to measure \Rd and \Rdst is required. The BaBar
measurement of the branching fraction of $\decay{\Bz}{\Dstarp \pim\pip\pim}$ reconstructs \Dstarp using the
$\decay{\Dstarp}{\Dz\pip}$ decay with $\decay{\Dz}{\Km\pip}$~\cite{TheBABAR:2016vzj}. By adding
$\decay{\Dz}{\Km\pip\pip\pim}$, it is expected that the uncertainty can be reduced by 50\% in 5 years, reducing the
total external systematic to 3.5\% in Run III and beyond. The predictions for the \belleTwo uncertainties are taken from
Ref.~\cite{BelleIIBook}. The values used are shown in Table~\ref{table:RdRdst_inputs} and the projection of the
future impact is shown in Fig.~\ref{fig:RdRdst} using the \texttt{GammaCombo} package~\cite{gammacombo}. This shows
the significance of the future world average by combining the uncertainties from the SM predictions with the predicted
uncertainties of the \belleTwo and \lhcb experiments using their final datasets (with 50\invab at \belleTwo and 50\invfb
at \lhcb). It is clear that if the central values remain the same then the statistical power of the \belleTwo and \lhcb
experiments will be more than sufficient to reach 5$\sigma$. An additional figure in~\ref{sec:appendix_figs}, Fig.~\ref{fig:RdRdstA},
compares the current world average with the current SM prediction, alongside the projections for \belleTwo and \lhcb.

\begin{table*}
  \caption{The SM prediction, world average and predictions of the relative uncertainty of the \lhcb and \belleTwo measurements of \Rd and \Rdst at 10\invfb, 22\invfb and 50\invfb
      and at 5\invab and 50\invab respectively. \lhcb is expected to measure \Rd in the upcoming years.}
  \begin{center}
  \resizebox{\textwidth}{!}{
    \begin{tabular}{lccccccccc}
    \hline
    Measurement     & SM                                        &  Current World                 & Current                        & \multicolumn{5}{c}{Projected Uncertainty } \\
                    & prediction                                &  ~~~~~Average~~~~~             & ~~~Uncertainty~~~              & \multicolumn{2}{c}{\belleTwo}   & \multicolumn{3}{c}{\lhcb}           \\
                    & \small{(Ref.~\cite{Bernlochner:2017jka})} &  \small{(Ref.~\cite{PDG2016})} &  \small{(Ref.~\cite{PDG2016})} &  5\invab &  50\invab &  8\invfb  &  22\invfb  &  50\invfb \\
      \hline
      \Rd           & $(0.299\pm 0.003)$                        & $(0.403\pm 0.040\pm 0.024)$    &  $11.6\%$                      & $5.6\%$   & $3.2\%$    & -       & -        & -                \\
      \Rdst         & $(0.257\pm 0.003)$                        & $(0.310\pm 0.015\pm 0.008)$    &  $5.5\%$                       & $3.2\%$   & $2.2\%$    & $3.6\%$   & $2.1\%$  & $1.6\%$            \\
      \hline
  \end{tabular}}
  \label{table:RdRdst_inputs}
  \end{center}
\end{table*}

\section{New physics in electroweak penguins}
\label{sec:LFU}

In this section, prospects for new physics searches in $\decay{\bquark}{\squark}$ transitions are studied under the SM
hypothesis as well as in several NP scenarios, with special attention given to present anomalies. The future projections
for \belleTwo are reported in Ref.~\cite{BelleIIBook}. The future uncertainties for LHCb have been symmetrised
where appropriate and are obtained as stated previously. However, the uncertainty on $f_s/f_d$ on the branching fraction of \Bstophigamma measured at
\lhcb~\cite{Aaij:2012ita} is assumed to be irreducible. Estimates of branching fraction ratios, $R(X)$, rely on extrapolations from the muonic branching fractions assuming the
same ratio of efficiencies between the electron and muon modes as has been observed in the analysis of
$R(K)$~\cite{Aaij:2014ora}.
For current measurements, correlations are taken into account when available. Most measurements will be dominated by the statistical uncertainty for the studied milestones, with only a few exceptions as e.g. for the differential branching fractions $d\BF/d\qsq$ of \BtoKmumu and \BtoKstmumu, where the dominant systematic uncertainties arise from the branching ratio of the respective normalisation channels, the form factor models and data-simulation differences. Hence, correlations between the systematic uncertainties are assumed to be negligible in this study.
The development of theoretical uncertainties is much harder to predict.
For quantities accessible to lattice QCD, the expected improvment in computing power allows to safely assume significant
improvements on the five to ten year time scale considered here.
In semi-leptonic decays, this concerns in particular the hadronic form factors. Even though current
lattice calculations of $B\to K^*$ form factors also face systematic uncertainties due to the finite
$K^*$ lifetime, a solution of this challenge is realistic in the near future \cite{Agadjanov:2016fbd}.
For $B\to K$ form factors, this problem is absent.
It thus seems realistic to assume a reduction of all form factor uncertainties by a factor of two by the time
of reaching milestone~II \cite{BelleIIBook} and we assume this in our numerics.
For the remaining uncertainties, in particular systematic uncertainties due to non-factorizable hadronic
contributions, we conservatively assume they will stay the same as at present,
even though data-driven methods might allow to reduce them in the future \cite{Bobeth:2017vxj, Blake:2017fyh}.

For \btosll and radiative \btosgamma transitions, the effective Hamiltonian can be expressed as
\begin{align} \mathcal{H}_{\rm eff} &= -\frac{4G_F}{\sqrt{2}} \lambda_t  \sum \limits_i ( C_i O_i   + C^{\prime}_i
O^{\prime}_i  ) + \rm{h. c.}, \end{align}
where $G_F$ is the Fermi constant and $\lambda_t = \Vtb \Vtss$ is a CKM factor. In a large class of new physics models,
the most important new physics effects in these transitions appear in the Wilson coefficients $C_i$ of the following
dimension-6 operators,
\begin{alignat}{2} %O_S &=  \frac{e}{16\pi^2}m_b (\bar{s} P_R b) (\bar{\ell} \ell),\\
%O^{\prime}_{S} &= \frac{e^2}{16\pi^2}m_b(P_R \bar{s} b) (\bar{\ell} \gamma_5 \ell), \\
%O_P &=  \frac{e}{16\pi^2}m_b (\bar{s} P_R b)(\bar{\ell} \ell),\\
%O^{\prime}_{P} &=  \frac{e^2}{16\pi^2}m_b(P_R \bar{s} b) (\bar{\ell} \gamma_5 \ell), \\
O_7 &= \frac{e}{16\pi^2}m_b(\bar{s} \sigma^{\mu\nu} P_R b) F_{\mu\nu}, \\
O^{\prime}_{7} &= \frac{e^2}{16\pi^2}m_b(\bar{s} \sigma^{\mu\nu} P_L b) F_{\mu\nu}, \\
O_9 &= \frac{e}{16\pi^2}(\bar{s} \gamma_{\mu} P_L b) (\bar{\ell} \gamma^{\mu} \ell), \\
O^{\prime}_{9} &= \frac{e^2}{16\pi^2}(\bar{s} \gamma_{\mu} P_R b) (\bar{\ell} \gamma^{\mu} \ell), \\
O_{10} &= \frac{e^2}{16\pi^2}(\bar{s} \gamma_{\mu} P_L b) (\bar{\ell} \gamma^{\mu}\gamma_5 \ell), \\
O^{\prime}_{10} &= \frac{e^2}{16\pi^2}(\bar{s} \gamma_{\mu} P_R b) (\bar{\ell}\gamma^{\mu}\gamma_5 \ell).
 \end{alignat}
In the following considerations, the effective Wilson coefficient $C^{\rm eff}_7$ (see e.g.~\cite{Blake:2016olu}) is
used instead of $C_7$ as this effective coefficient is independent of the regularisation scheme, where we define

\begin{align} C^{\rm eff}_7 &= C^{\rm eff \, SM}_7 + C^{\rm NP}_7,\\ C^{\prime \, \rm eff}_7 &= C^{\prime \, \rm eff \,
SM}_7 + C^{\prime \, \rm NP}_7. \end{align}

The impact of future measurements is studied by performing scans of the new physics contribution to the Wilson coefficients at a scale of $\mu=4.8\gev$,
using the \texttt{flavio}~\cite{flavio} package, under the SM hypothesis and several different new physics scenarios, listed in Table~\ref{tab:NPmodels}.
The measurements are separated depending on whether they are inclusive or exclusive. This allows
for a proper comparison given their respective uncertainties have different origins. Various NP scenarios are chosen for each class of measurement and each scan parameter
on the basis of existing global fits~\cite{Hurth:2016fbr,Paul:2016urs,Capdevila:2017bsm,Altmannshofer:2017fio,Altmannshofer:2017yso,Geng:2017svp}.
Scans to $C_S$ and $C_P$ (see e.g.~\cite{Blake:2016olu}) are omitted as these are dominated by contributions from purely leptonic $\decay{\B}{\ellp\ellm}$ decays, where, apart from for \Bsmumu, only limits are available as indicated in Table~\ref{tab:limits}.\\

\begin{table*}
\caption{New physics scenarios for \lhcb, \belleTwo exclusive and \belleTwo inclusive Wilson coefficient scans. Contributions to the Wilson coefficients arising from new physics are given for each scan.}
\begin{center}
  \resizebox{\textwidth}{!}{
    \begin{tabular}{lccccc}
    \hline
     & $(\Cnine, \Cten)$  & $(\Cninep, \Ctenp)$ & $(\Cnine, \Cnineee)$ & $(\ReCsevp, \ImCsevp)$ & $(\ReCsev, \ImCsev)$ \\
       \hline
       \lhcb & $(-1.0, 0.0)$ & $(-0.2, -0.2)$ & $(-1.0, 0.0)$ & $(0.00, 0.04)$ & $(-0.075, 0.000)$\\
       \belleTwo exclusive & $(-1.4, 0.4)$ & $(0.4, 0.2)$ & $(-1.4, -0.7)$ & $(0.08, 0.00)$ & $(-0.050, 0.050)$\\
       \belleTwo inclusive & $(-0.8, 0.6)$ & $(0.8, 0.2)$ & $(-0.8, 0.4)$ & $(0.02, -0.06)$ & $(-0.050, -0.075)$\\
                   \hline
  \end{tabular}
  }
\end{center}
\label{tab:NPmodels}
\end{table*}

The scans of the electromagnetic dipole coefficients $C^{(\prime)}_7$ rely on measurements of the branching fractions of \Bstophigamma, \BtoKstpgamma, \BtoKstzgamma, \BtoXsgamma, on $\ADeltaG(\Bstophigamma)$ and \SKstgamma as well as \ATtwo (also known as $P_1$) and \ATIm extracted from \BtoKstee decays at very low $\qsq$. Furthermore, the angular observables $A_{7,8,9}$ in \BtoKstmumu constrain the imaginary part of $C^{(\prime)}_7$.

The measurements entering the scans of the semi-leptonic coefficients $C^{(\prime)}_{9,10}$ comprise the inclusive $\BF(\BXsmumu)$ at low and high $\qsq$; the low $\qsq$ range is split equally for extrapolations. The forward-backward asymmetry $A_\text{FB}(\BXsll)$ has been measured at low and high $\qsq$, and extrapolations to future sensitivities are available in several low and high $\qsq$ ranges. The differential branching fractions $d\BF/d\qsq$ of \BtoKmumu, \BtoKstmumu and \Bstophimumu decays in both low and high $\qsq$ regions is included in the scans, as well as the angular observables $S_{3,4,5}, F_L, A_\text{FB}$ in several bins of $\qsq$ from \lhcb. The angular observables available for Belle~(II) are $P^{\prime}_{4,5}(\BtoKstmumu)$ in similar ranges. Scans of $C^{(\prime)}_{10}$ further include the branching fraction of the decay \Bsmumu.

In the scan of \Cnine vs. \Cnineee, $P^{\prime}_{4,5}$ extracted from
\BtoKstee decays is included in addition to the muonic final
state. Information on electrons is further obtained from the ratios of
branching fraction between muon and electron final states for $R(X_s)$,
$\Rk$, $\Rkst$ and $R(\phi)$. The results of the \belle collaboration on \Rk and \Rkst in the region $0.0  < \qsq < 22.0 \gev^2$ were not considered as input in this scan as the charmonium region is included~\cite{Wei:2009zv}. The inclusive measurement of $R(X_s)$ will become accessible at \belleTwo, whereas $R(\phi)$ will be measurable at \lhcb at low and high $\qsq$.
Measurements of lepton flavour universality pose stringent tests on the SM and several tensions have already been observed as mentioned briefly in the previous section. The \lhcb collaboration found \Rk to be $0.745^{+0.090}_{-0.074} \pm 0.036$~\cite{Aaij:2014ora}; $2.6\sigma$ below the SM expectation. The symmetrised uncertainty on \Rk in $1.0  < \qsq < 6.0 \gev^2$ is expected to be $0.046$ at milestone I, $0.025$ at milestone II and down to $0.016$ by milestone III. The uncertainties in the range $15.0  < \qsq < 22.0 \gev^2$ are expected to behave similarly. A recent measurement of \Rkst by the \lhcb collaboration~\cite{Aaij:2017vbb} finds a tension of $2.1-2.3\sigma$ in $0.045  < \qsq < 1.1 \gev^2$ and of $2.4-2.5\sigma$ in $1.1  < \qsq < 6.0 \gev^2$ with respect to the available SM predictions. The measured values of \Rkst are $0.66^{+0.11}_{-0.07} \pm 0.03$ and $0.69^{+0.11}_{-0.07} \pm 0.05$ in the ``very low" and "low" $\qsq$ regions respectively~\cite{Aaij:2017vbb}. The symmetrised uncertainties are extrapolated to future datasets and expected to be $0.048\,(0.053)$, $0.026\,(0.028)$ and $0.017\,(0.019)$ after milestones I, II and III, respectively, for low (central) $\qsq$ regions. Both the \Rk and \Rkst measurements of \lhcb will be dominated by the statistical uncertainty for all considered future milestones, wherefore correlations between the various systematic uncertainties can be neglected. If the anomalies in \Rk and \Rkst persist at the current central values, \lhcb will measure \Rk with a significance of $>5\sigma$ with respect to the SM prediction at milestone I, increasing to $15\sigma$ with the milestone III dataset. Concerning \Rkst at low $\qsq$, the tension would increase to $3.4-3.8\sigma \, (6.2-6.9\sigma)$, depending on the SM prediction, at milestone I (II); a tension of around $10\sigma$ would be reached by milestone III. For \Rkst at high $\qsq$, a tension of $4.7-4.8\sigma$ would emerge when reaching milestone I increasing to $9.0-9.1\sigma \, (13.2-13.4\sigma)$ at milestone II (III). If the anomalies in \btosll decays persist, the \belleTwo collaboration will be able to confirm the anomalies in \Rk (\Rkst) when reaching the integrated luminosity of milestone II in the region $1.0 \, (1.1)  < \qsq < 6.0 \gev^2$ with significances around $7-8\sigma$ and hence tensions of this size will be conclusively observed within the next few years.

\begin{figure*}
    \centering
    \subfigure[$\;$\Cnine versus \Cten.]{
      \includegraphics[width=0.46\textwidth]{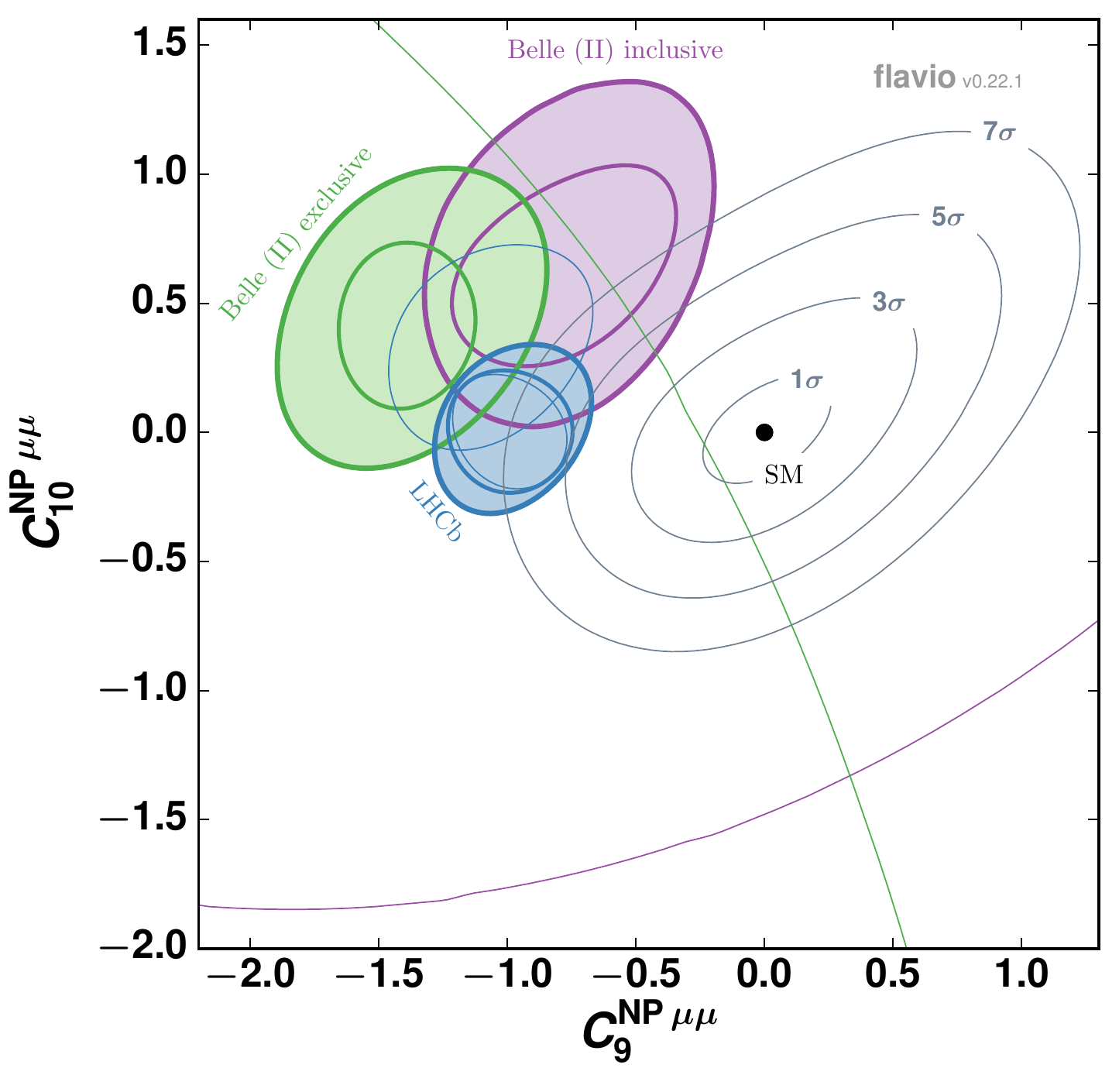}
      \label{fig:C9C10}
    }
    \subfigure[$\;$\Cnine versus \Cnineee.]{
      \includegraphics[width=0.46\textwidth]{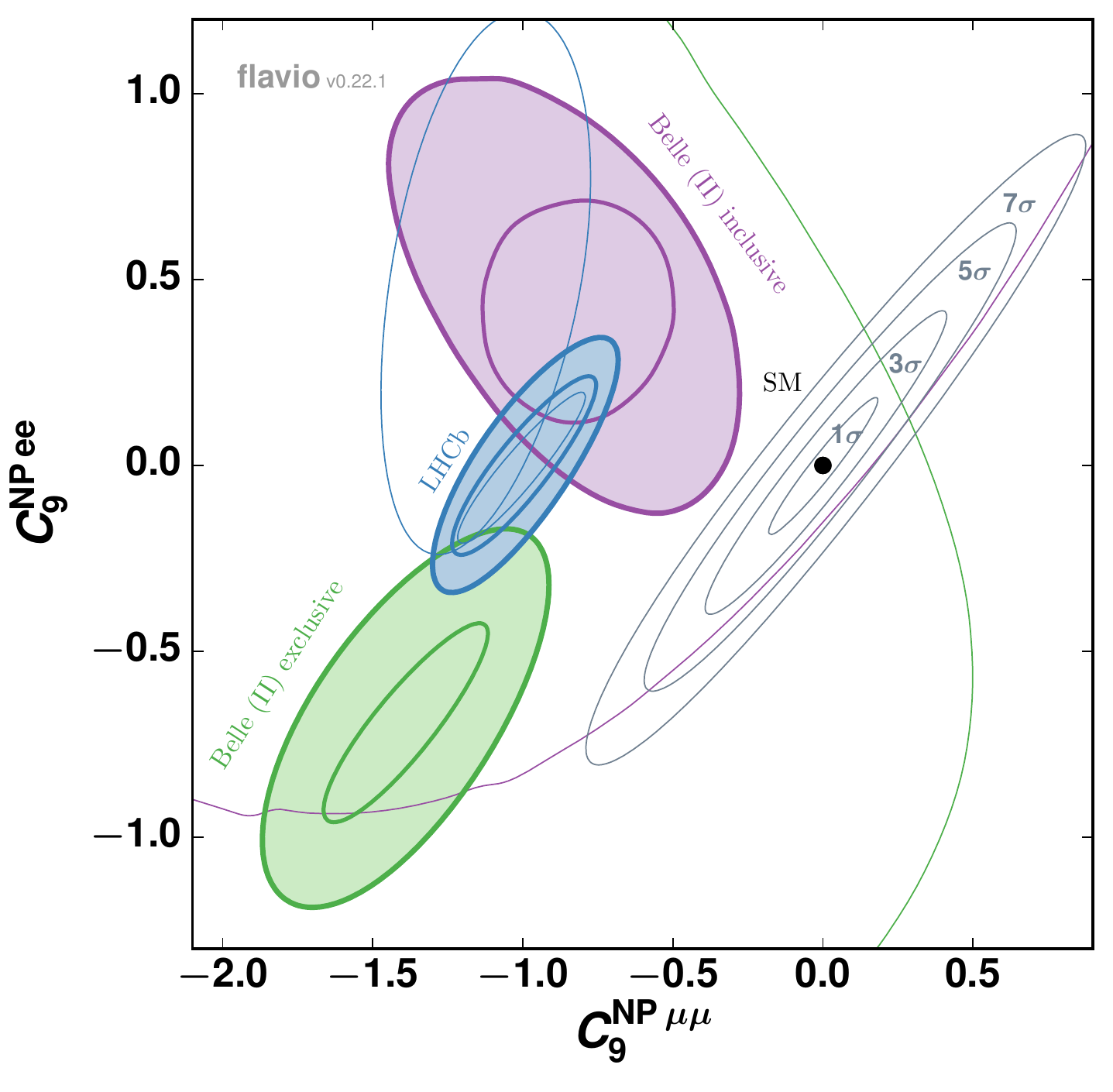}
      \label{fig:C9mumuC9ee}
    }
    \subfigure[$\;$\ReCsev versus \ImCsev.]{
      \includegraphics[width=0.46\textwidth]{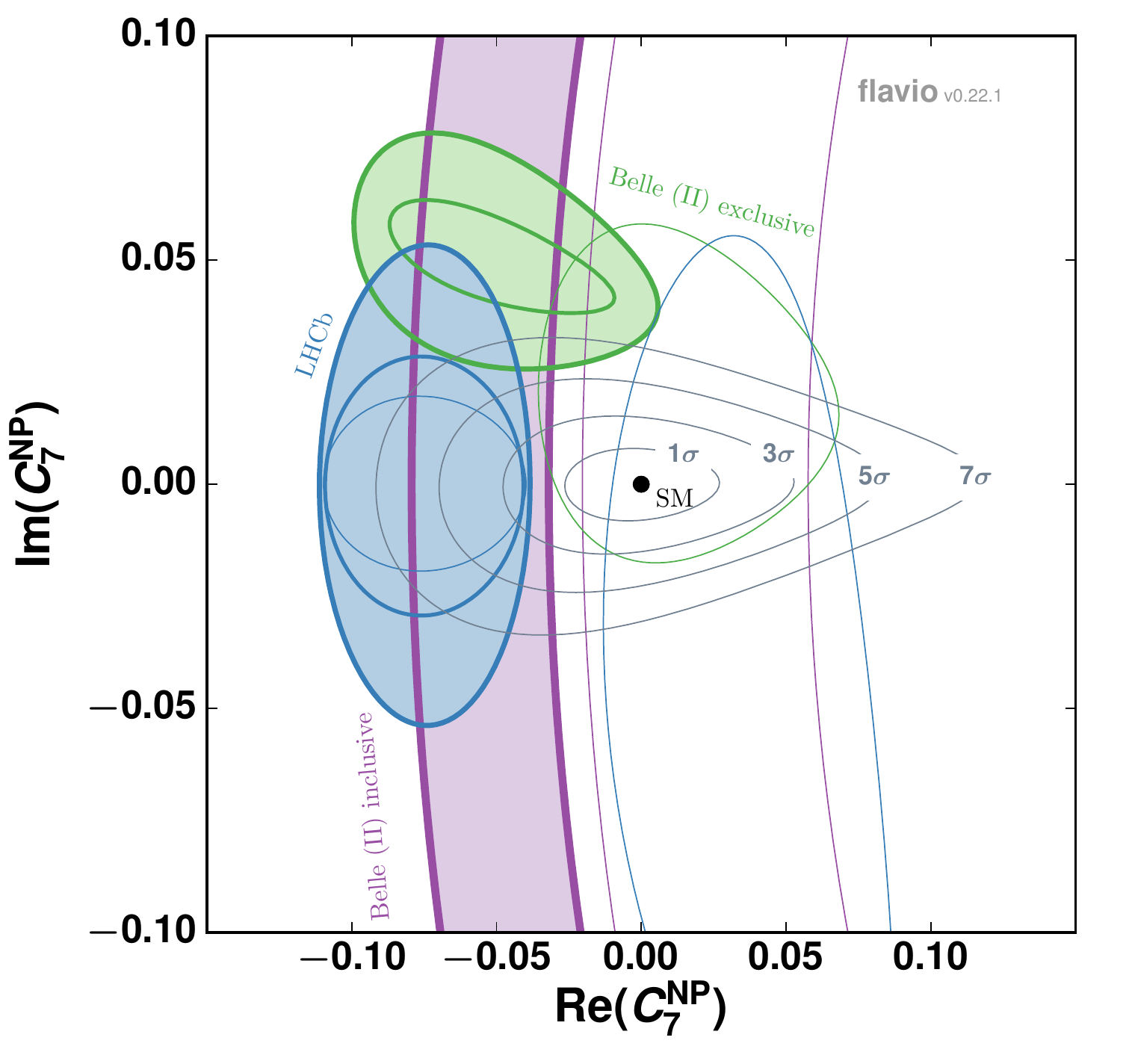}
      \label{fig:C7}
    }
    \caption{In the two-dimensional scans of pairs of Wilson coefficients, the current average (not filled) as well as the extrapolations to future sensitivities (filled) of \lhcb at milestones I, II and III (exclusive) and \belleTwo at milestones I and II (inclusive and exclusive) are given. The central values of the extrapolations have been evaluated in the NP scenarios listed in Table~\ref{tab:NPmodels}. The contours correspond to $1\sigma$ uncertainty bands. The Standard Model point (black dot) with the $1\sigma, 3\sigma$, $5\sigma$ and $7\sigma$ exclusion contours with a combined sensitivity of \lhcb's $50\invfb$ and \belleTwo's $50\invab$ datasets is indicated in light grey. The primed operators show no tensions with respect to the SM; hence no SM exclusions are provided. \\
    \mbox{} \\ % to get rid of single hanging line on one page
    \mbox{} \\
    }
    \label{fig:WC}
\end{figure*}

The scans of the unprimed semi-leptonic and electromagnetic dipole Wilson coefficients are illustrated in Fig.~\ref{fig:WC}, where detailed information on the chosen inputs together with the scans of the primed operators are given in~\ref{sec:appendix_LFU}. As illustrated in Figs.~\ref{fig:C9pC10p} and~\ref{fig:C7p}, no discrepancies to the SM for the primed operators is visible.
The electromagnetic dipole operators are currently consistent with the SM hypothesis and the contours obtained from \lhcb, inclusive and exclusive \belleTwo measurements are in good agreement.
The current measurements hint at a deviation from the SM in the unprimed operator \Cnine, which prefers a negative value driven by the \lhcb measurements. In contrast to the tension observed in \mbox{\Cnine,} no hints towards new physics are visible in \Cnineee, nor in \Cten.
%If the current anomaly in \Cnine persists, the combined sensitivity of \lhcb and \belleTwo at milestone III will
%allow for clarification of whether there is a new physics contribution to \Cnine.
Even if the curent tensions seen in \bsll data turn out to be statistical fluctuations,
there are many very rare decays, lepton flavour violating decays,
and decays with neutrinos in the final state that are orthogonal clean probes of NP
(see e.g.\ \cite{Buras:2014fpa,Glashow:2014iga,Fleischer:2017ltw,Calibbi:2017uvl}).
Corresponding sensitivities are listed in Table~\ref{tab:limits}. For
the determination of the sensitivity of \Bstautau, the conservative
assumption of the same trigger improvement as for a decay with a single tau lepton was used. The extrapolations of \Bsee are extracted from the latest \lhcb measurement~\cite{Aaij:2017vad} of \Bsmumu by factoring in an electron penalty factor. Following the approach in~\cite{Aaij:2013fia} for the lepton-flavour violating decay \tautothreemu, the $\tau$ production cross section was scaled linearly with the centre-of-mass energy.

\begin{table*}
\caption{Expected sensitivities of specific very rare decays; limits are given at 90\% \mbox{C. L. .} Note that \belleTwo has sensitivity for $\Bsll$, but we only consider the impact of the
$e^+e^- \to \Upsilon(4S) \to B \bar B$ data taking in this study. The extrapolations of \Bsmumu refer to the combined statistical and systematic uncertainty and are based on the latest \lhcb measurement on a dataset corresponding to an integrated luminosity of $4.4\invfb$~\cite{Aaij:2017vad}.}
\begin{center}
  \begin{threeparttable}
  \resizebox{\linewidth}{!}{
  \begin{tabular}{llccccc}
    \hline
     & current & \multicolumn{3}{c}{\lhcb} & \belleTwo \\
     & & $8 \invfb$ & $22 \invfb$ & $50 \invfb$ & $50\invab$  \\
           \hline
           \Bsmumu  & $(2.4^{+0.9}_{-0.7})\times 10^{-9}$~\cite{PDG2016}\tnote{\hyperlink{fnthree}{iii}} & $0.45 \times 10^{-9}$ & $0.24 \times 10^{-9}$ & $0.16 \times10^{-9}$ &  - \\
           \Bdmumu & $< 0.28 \times 10^{-9}$~\cite{Aaij:2017vad}\tnote{\hyperlink{fnfour}{iv}}  & $< 0.19 \times 10^{-9}$ & $< 0.10 \times 10^{-9}$ & $< 0.07 \times 10^{-9}$ & $< 5 \times 10^{-9}$  \\
	  \Bsee & $< 2.8 \times 10^{-7}$~\cite{Aaltonen:2009vr} & $< 0.27 \times 10^{-8}$ & $<0.12 \times 10^{-8}$ & $<0.07 \times10^{-8}$ & - \\
	  \Bdee & $< 8.3 \times 10^{-8}$~\cite{Aaltonen:2009vr} & $< 0.12 \times 10^{-8}$ & $<0.05 \times 10^{-8}$ & $<0.03 \times10^{-8}$ & $<3 \times10^{-9}$  \\
	  \Bstautau & $ <5.2\times 10^{-3}$~\cite{Aaij:2017xqt} & $ <2.7\times 10^{-3}$ & $<0.9\times 10^{-3}$ & $<0.5\times 10^{-3}$ & -  \\
	 \Bdtautau & $ < 1.6\times 10^{-3}$~\cite{Aaij:2017xqt} & $ < 0.8\times 10^{-3}$ & $ < 0.3\times 10^{-3}$ & $ < 0.2\times 10^{-3}$ & $<0.3\times 10^{-3}$  \\
	\Bsemu  & $<1.1 \times 10^{-8}$~\cite{Aaij:2013cby}\tnote{\hyperlink{fnfive}{v}}  & $<0.31 \times 10^{-8}$ & $<0.15 \times 10^{-8}$ & $<0.10 \times 10^{-8}$ & -  \\
	\Bdemu  & $<2.8 \times 10^{-9}$~\cite{Aaij:2013cby}\tnote{\hyperlink{fnfive}{v}}  & $<0.8 \times 10^{-9}$ & $<0.4 \times 10^{-9}$ & $<0.2 \times 10^{-9}$ & $< 4.0 \times 10^{-9}$  \\
\tautothreemu  & $<2.1  \times 10^{-8}$~\cite{Hayasaka:2010np} & $<2.4 \times 10^{-8}$~\cite{Aaij:2013fia} & $<1.3 \times 10^{-8}$ & $<0.8 \times 10^{-8}$ &  $<3.5 \times 10^{-10}$\\
\tautomugamma  &  $<4.4 \times 10^{-8}$~\cite{Aubert:2009ag} & - & - & - &  $<1.0 \times 10^{-9}$ \\
\BtoKpnunu  &  $<1.6 \times 10^{-5}$~\cite{Lees:2013kla} & - & - & - & $10.7\%$~\cite{Knunu}\\
\BtoKstpnunu  & $<4.0 \times 10^{-5}$~\cite{Lutz:2013ftz} & - & - & - & $9.3\%$~\cite{Knunu}\\
\BtoKstznunu  & $<5.5 \times 10^{-5}$~\cite{Lutz:2013ftz} & - & - & - & $9.6\%$~\cite{Knunu}\\
                   \hline
  \end{tabular}}
  \begin{tablenotes}
    \hypertarget{fnthree}{\item[iii]{This average does not contain the latest \lhcb measurement~\cite{Aaij:2017vad}.}}
    \hypertarget{fnfour}{\item[iv]{From supplementary material. A combination of measurements is available from~\cite{PDG2016}.}}
  \hypertarget{fnfive}{\item[v]{This measurement has been performed on 1 \invfb and has been extrapolated to 3\invfb.}}
  \end{tablenotes}
  \end{threeparttable}
  \end{center}
\label{tab:limits}
\end{table*}

\section{Conclusion}
\label{sec:conclusion}

Projections of the future sensitivity of the \belleTwo and \lhcb datasets
have been analysed with regard to several important flavour physics observables.
For the first time, the complimentarity and combination of the two experiments
has been studied. Sensitivty estimates and projections have been made for several
important future milestones, corresponding to an intermediate point in \belleTwo and \lhcb
data taking (2020), the end of \belleTwo data taking (2024) and the end of scheduled \lhcb data taking (2029).
The foreseen changes in the trigger system of \lhcb are considered as well
as the anticipated scaling of the systematic uncertainties at both experiments.
This manuscript focuses on present day anomalies and other key measurements in the flavour sector, such as the
CKM angle $\gamma$ will be measured with a precision below 1$^\circ$.
There has been a long standing discrepancy between
the inclusive and exclusive determination of $\Vub$ (and to some extent also $\Vcb$), which
will, if the current central values remain, be established with a
significance well beyond 5$\sigma$. Further tensions have been observed in tests of lepton flavour universality in tree-level and loop-level processes. The current HFLAV average of the ratio of ${\decay{\B}{D^{(*)} \ell \neu}}$ tree-level decays involving $\tau$ leptons and light leptons, \Rd and \Rdst, differs from the Standard Model prediction by 3.9$\sigma$. The future
measurements will yield precisions of 3.2\% and 1.3\%, for \Rd and
\Rdst respectively (which does not include the potential for \lhcb to also measure \Rd). If the current central values persist, the
SM prediction can be ruled out by the combined dataset of
\belleTwo and \lhcb with a significance of well beyond 10$\sigma$. Further hints at a possible violation of lepton flavour universality have emerged in flavour-changing neutral current decays based on $\decay{\bquark}{\squark}$ transitions, which are a sensitive probe of new physics. The reach of
the future experiments is analysed by a scan of the Wilson coefficients
under different new physics scenarios and the SM
hypothesis. Currently, a set of anomalies in a range of observables,
from lepton flavour universality to branching ratio and angular
observables, are seen with local significances ranging between $2.5 -
3.9\sigma$.
The current combination of these anomalies is reported by several groups performing global fits, some of which quote
deviations with a significance well beyond $5\sigma$. However, depending on the treatment of hadronic
uncertainties, the significance can be considerably less. More data and more work on the theoretical side are needed to
clarify the situation.
% is on the edge
% of being significant at the level of 5$\sigma$ as reported by several
% groups performing global fits.
If the anomalies in \btosll decays persist, a highly significant
tension will be observed within the next years by the two single
experiments independently.

Even though both the \belleTwo and the \lhcb experiments could
individually confirm or rule out many of the current flavour anomalies, the
advantage of the experimental situation in the near future is that a potential anomaly can
be cross-checked by a competing experiment, with mostly
orthogonal systematic uncertainties. Furthermore, the two experiments are designed in very
different ways and subsequently have sensitivity to different regions of phase space, where comparison
between them will be vital to establish evidence of new physics.
For example \belleTwo has better sensitivity for inclusive measurements and those involving neutral final states, whereas
\lhcb has typically better performance for exclusive modes or those involving very rare decays.

We conclude that if the current flavour anomalies persist, both \belleTwo and \lhcb will have overwhelming evidence for new physics
on a short time scale. If these deviations are hints of physics beyond the SM, combining information from both experiments will be vital
to understand its nature. Furthermore, unprecendeted advances in tree-level CKM parameter precision measurements can be expected in the near future.
Comparison of these values with the loop-level counterparts will scrutinize the unitarity of the SM and probe new physics at high energy scales.

\pagebreak
\section{Acknowledgements}

We would like to thank Marcello Rotondo, Stefan Schacht and Robert Fleischer for their kind feedback
that helped to improve the manuscript. J. A. gratefully acknowledges support of the Deutsche
Forschungsgemeinschaft (DFG, Emmy Noether programme: AL 1639/1-1)
and of the European Research Council (ERC Starting Grant: PRECISION 714536).
F. B. gratefully acknowledges support of the Deutsche Forschungsgemeinschaft (DFG, Emmy Noether programme: BE 6075/1-1).
The work of D.~S. was supported by the DFG cluster of
excellence ``Origin and Structure of the Universe''.
D.~S. thanks Javier Virto and Tobias Huber for fruitful discussions.
M.~K. and A.~T. would like to
acknowledge support from the UK national funding agency, STFC, and Clare College,
University of Cambridge.

\clearpage
\bibliographystyle{elsarticle-num}
\bibliography{paper}

\appendix
\clearpage
\section{Additional Figures}
\label{sec:appendix_figs}

In this appendix some additional figures are provided. These use no additional information to those shown in the main text but simply offer a different perspective.

\begin{figure}[h]
  \centering
  \includegraphics[width=0.7\textwidth]{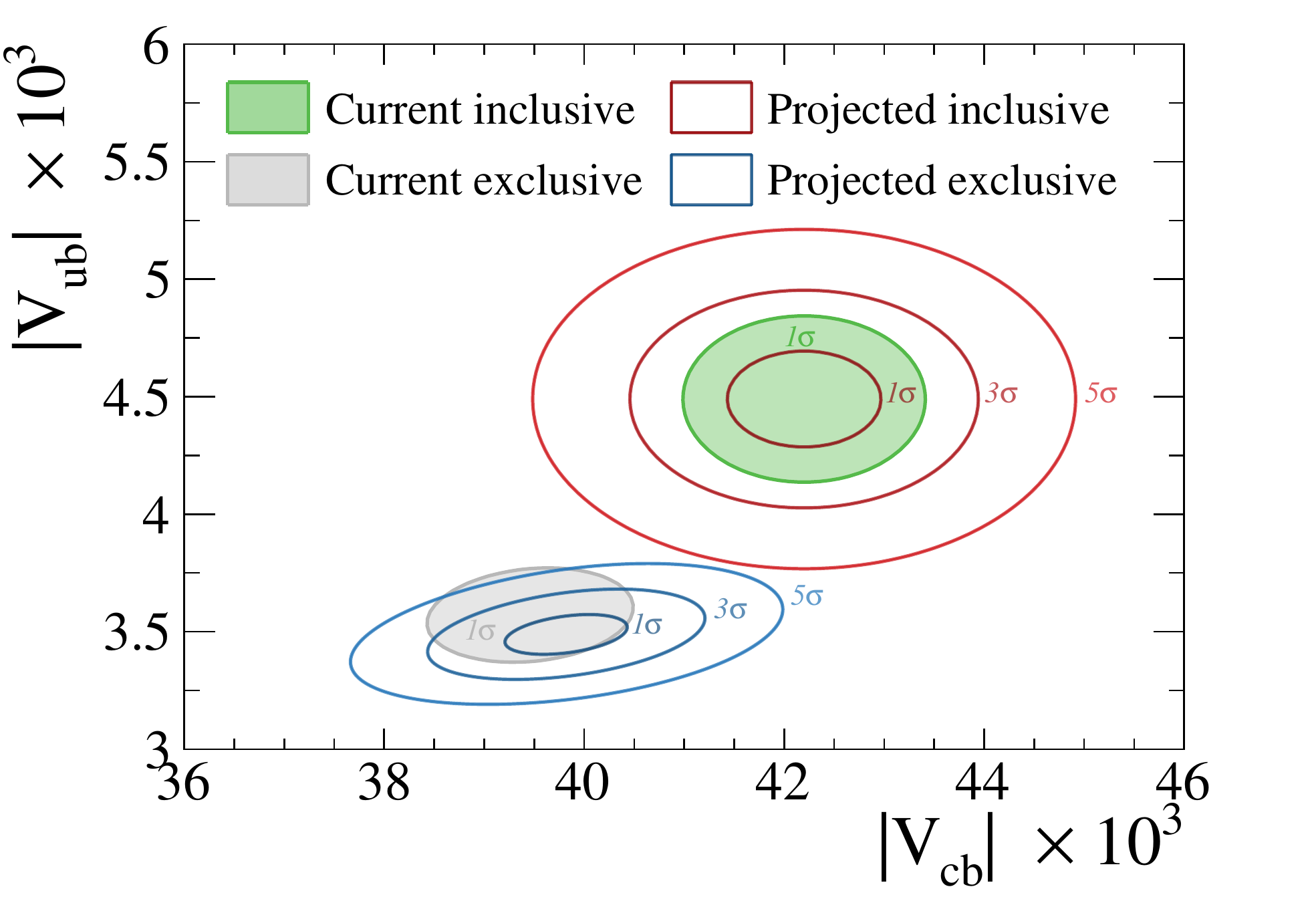}
  \caption{The current (filled areas) and future compatibility, at milestone III, (non-filled areas) of inclusive and exclusive methods for measurements of \Vub and \Vcb}
  \label{fig:VubVcbB}
\end{figure}

\begin{figure}[h]
  \centering
  \includegraphics[width=0.7\textwidth]{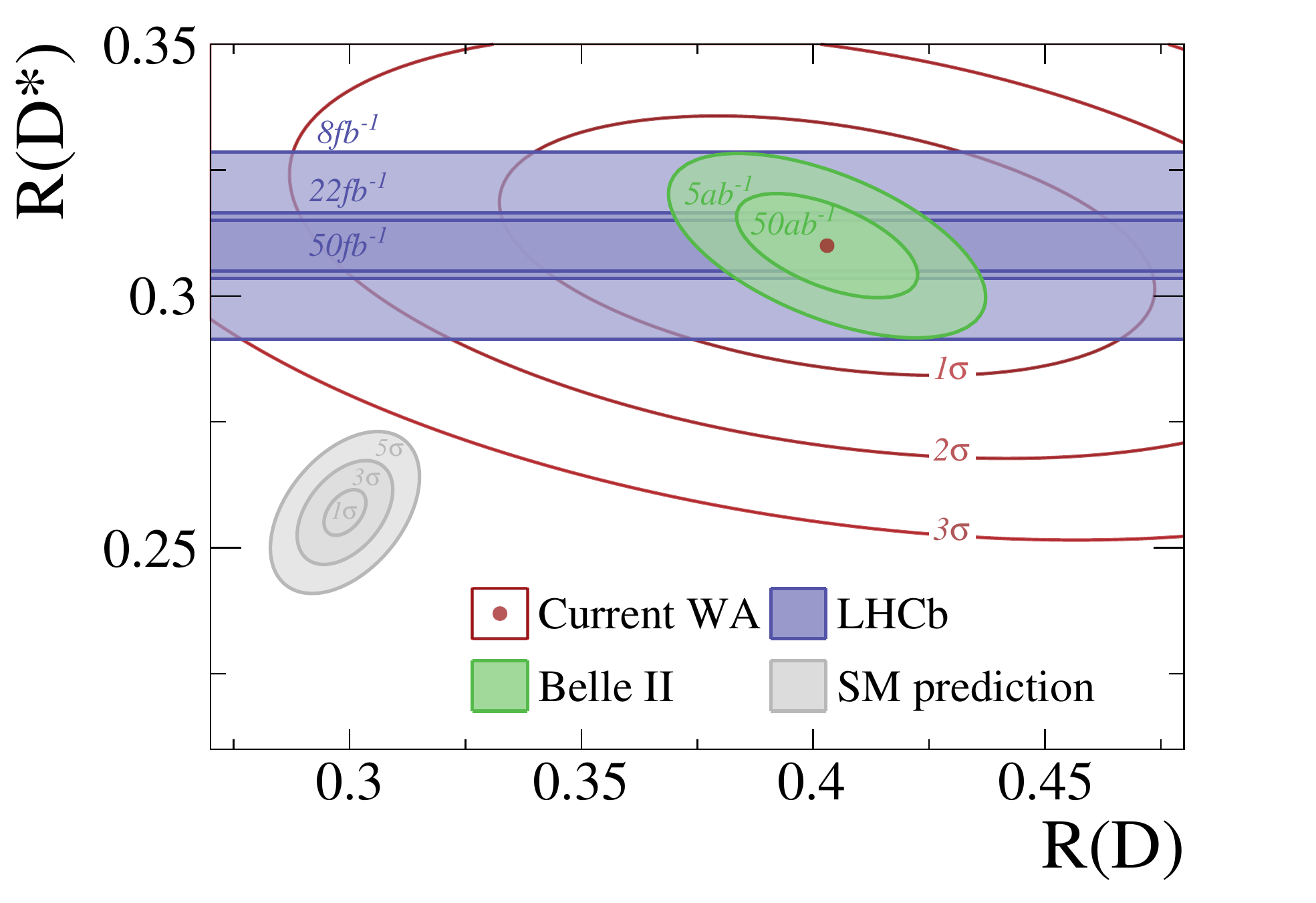}
  \caption{The current world average (red lines), the SM prediction (gray bands) and future prospects from \belleTwo (green) and \lhcb (blue) for measurements of \Rd and \Rdst.}
  \label{fig:RdRdstA}
\end{figure}

\begin{figure}[h]
  \centering
  \includegraphics[width=0.7\textwidth]{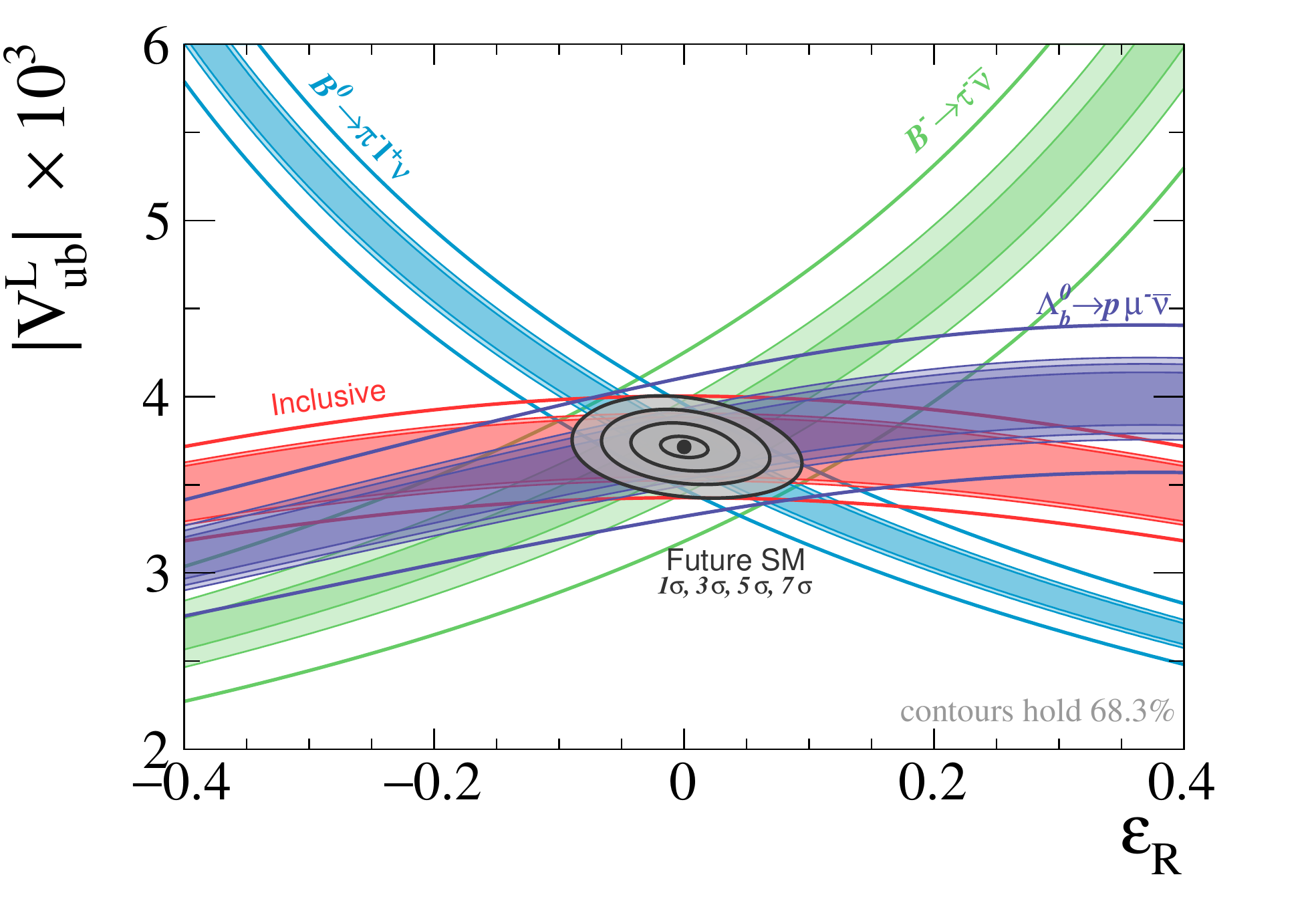}
  \caption{Prospects for new physics measurements related to right handed currents with the central values shifted to the SM expectation. The current uncertainties (not filled) are shown
    along with the future projections at milestones I, II and III (filled). The SM point (black dot) with the $1\sigma$, $3\sigma$, $5\sigma$ and $7\sigma$ exclusion contours (black lines)
    at milestone III are overlaid. Note that changing the central value of \Vub used in this case
  simply serves to shift the $y$-axis.}
  \label{fig:rh_currents_central}
\end{figure}

\begin{figure}[h]
  \centering
  \includegraphics[width=0.7\textwidth]{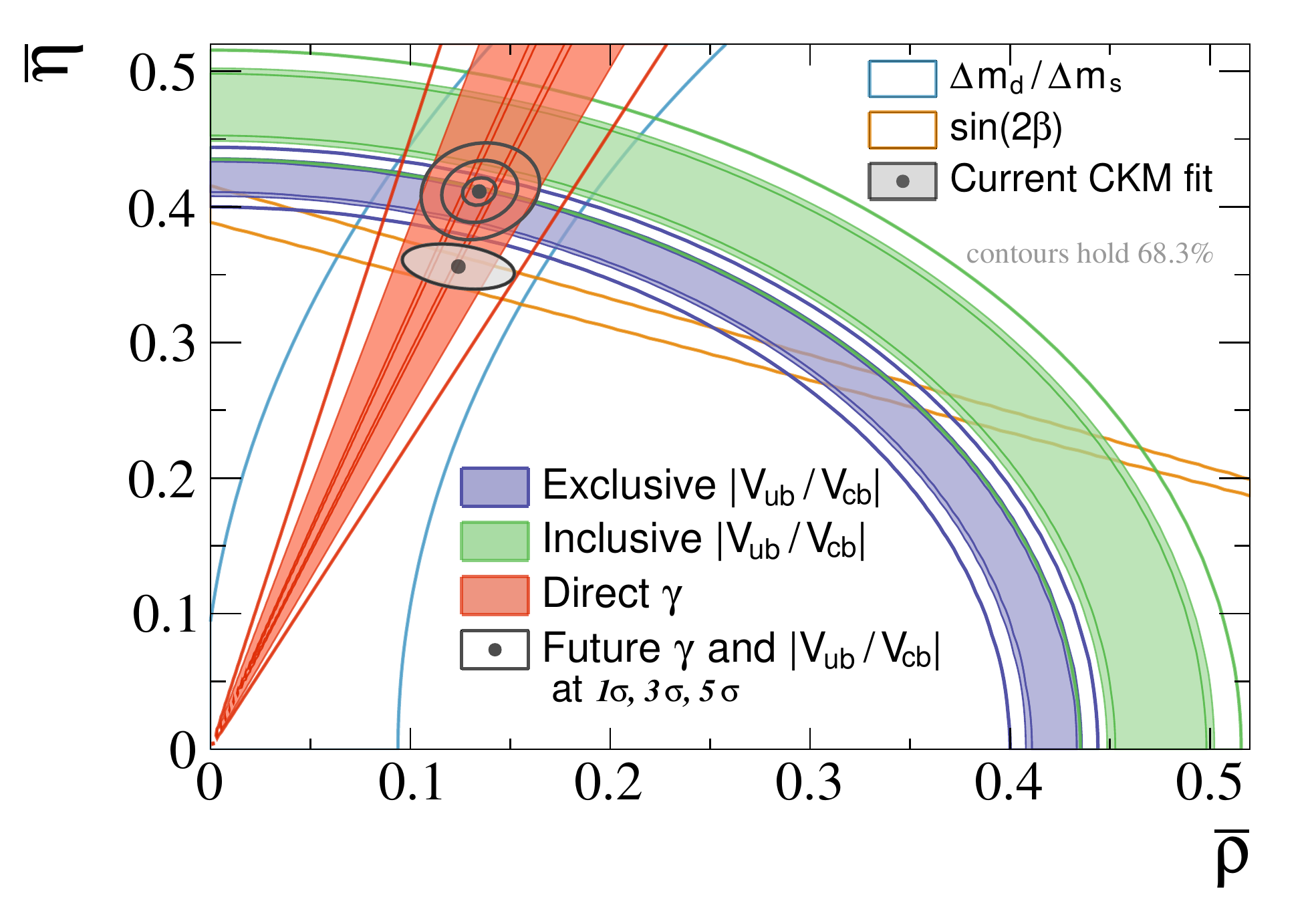}
  \caption{Prospects for CKM fits in $(\rhobar,\etabar)$ space. This is the same plot as shown in Fig.~\ref{fig:RhoEta} with the current loop-level constraints from $\sin(2\beta)$, $\Delta m_{d}$ and $\Delta m_{s}$ (values from Ref.~\cite{PDG2016}) overlaid.}
  \label{fig:RhoEta2}
\end{figure}

\begin{figure*}
    \centering
    \subfigure[$\;$\Cninep versus \Ctenp.]{
      \includegraphics[width=0.43\textwidth]{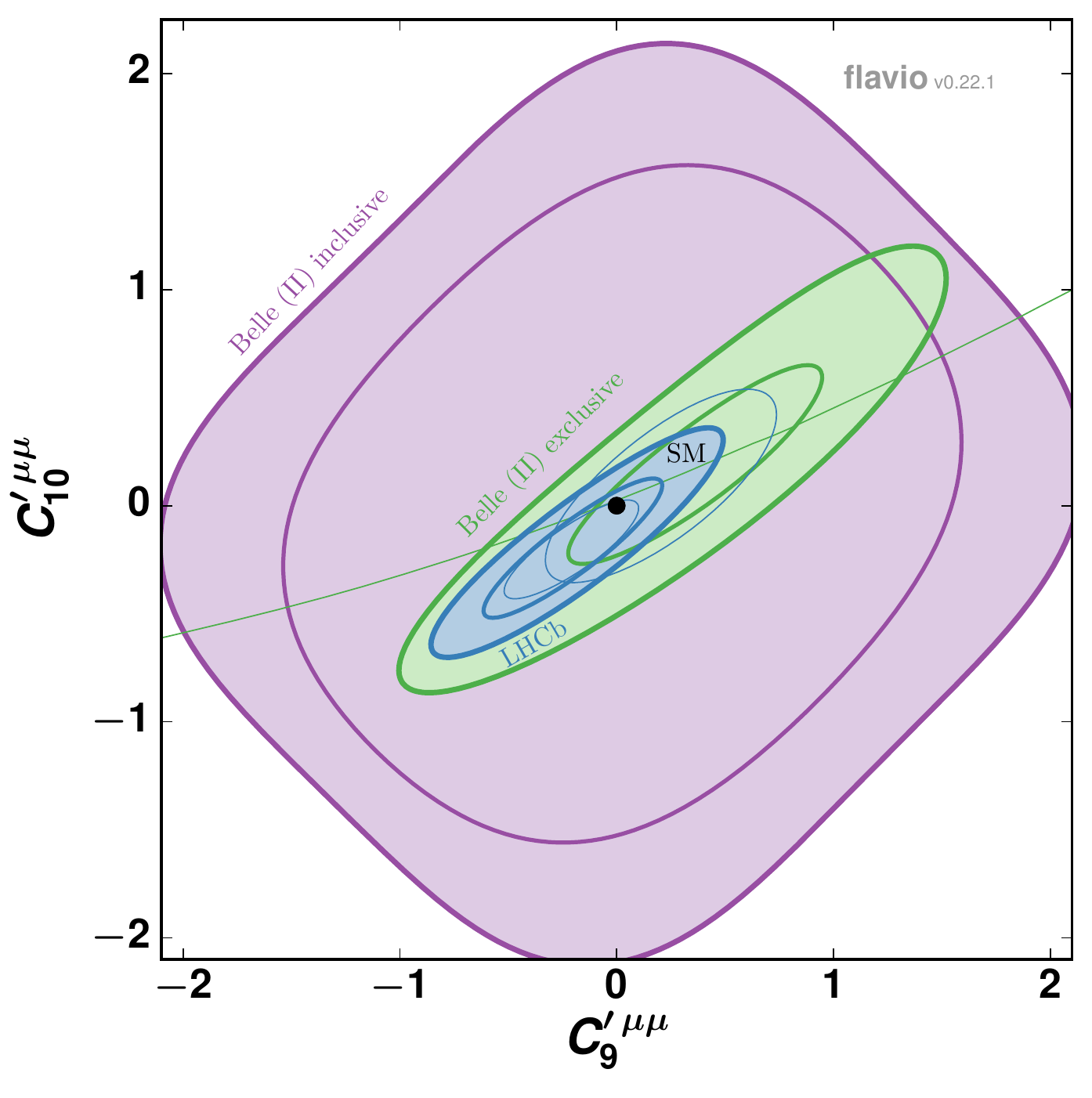}
      \label{fig:C9pC10p}
    }
    \subfigure[$\;$\ReCsevp versus \ImCsevp.]{
      \includegraphics[width=0.45\textwidth]{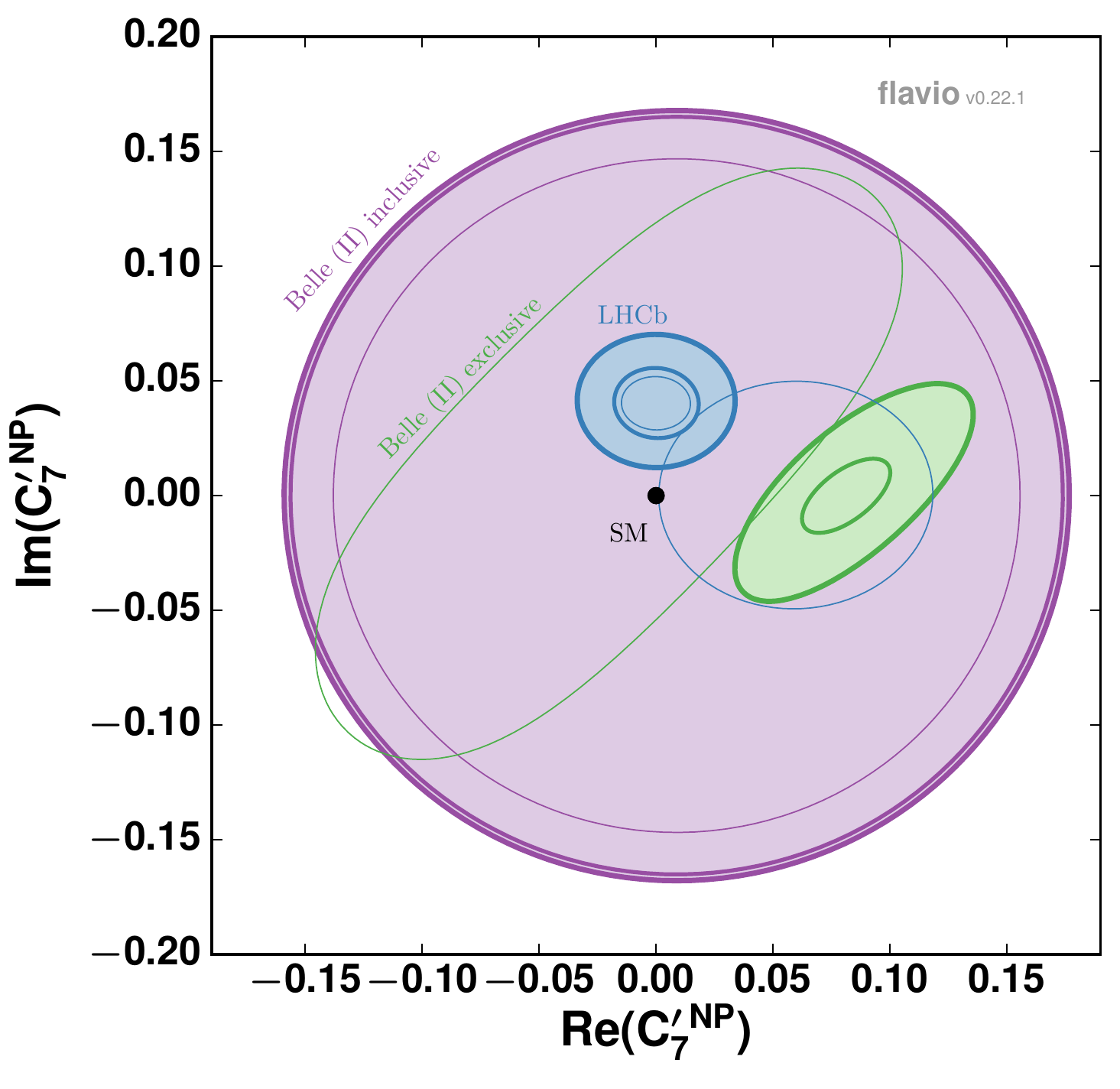}
      \label{fig:C7p}
    }    \caption{In the two-dimensional scans of pairs of Wilson coefficients, the current average (not filled) as well as the extrapolations to future sensitivities (filled) of \lhcb at milestones I, II and III (exclusive) and \belleTwo at milestones I and II (inclusive and exclusive) are given. The central values of the extrapolations have been evaluated in the NP scenarios listed in Table~\ref{tab:NPmodels}. The contours correspond to $1\sigma$ uncertainty bands. The Standard Model point (black dot) with the $1\sigma, 3\sigma$, $5\sigma$ and $7\sigma$ exclusion contours with a combined sensitivity of \lhcb's $50\invfb$ and \belleTwo's $50\invab$ datasets is indicated in light grey. The primed operators show no tensions with respect to the SM; hence no SM exclusions are provided.}
    \label{fig:WC_app}
\end{figure*}

\clearpage
\section{Inputs to Wilson coefficient scans}
\label{sec:appendix_LFU}

Details on the observables included in the scans of the semi-leptonic and electromagnetic dipole Wilson coefficients are given in Tables~\cref{tab:inputsSL,tab:inputsSLspecifics,tab:inputsRad}. The Belle (II) measurements entering the current average and extrapolated sensitivities to $5\invab$ and $50\invab$ are summarised in Tables~\cref{tab:extrapolationsBelleincl,tab:extrapolationsBelleexcl1,tab:extrapolationsBelleexcl2}, whereas the \lhcb measurements are detailed in Tables~\cref{tab:extrapolationsLHCb1,tab:extrapolationsLHCb2}. Corresponding central values as obtained with \texttt{flavio} for the SM and various NP models are given in Tables~\cref{tab:NPBelleincl,tab:NPBelleexcl1,tab:NPBelleexcl2,tab:NPLHCb1,tab:NPLHCb2}. The results of the scans to primed operators are illustrated in Fig.~\ref{fig:WC_app}.

\begin{table*}
\caption{Common inputs for the scans of the semi-leptonic Wilson coefficients. It is indicated if the observable is included in the current average and/or the extrapolations to future milestones.}
\begin{center}\begin{tabular}{lccc}
    \hline
    Observable & $\qsq$ interval & current average & extrapolations \\
    \hline
    $\BF(\BXsmumu)$ & $1.0  < \qsq < 3.5 \gev^2$ & - & \checkmark \\
    $\BF(\BXsmumu)$ & $3.5  < \qsq < 6.0 \gev^2$ & - & \checkmark \\
    $\BF(\BXsmumu)$ & $1.0  < \qsq < 6.0 \gev^2$ & \checkmark & - \\
    $\BF(\BXsmumu)$ & $\qsq  > 14.4 \gev^2$ & \checkmark & \checkmark \\
    $A_\text{FB}(\BXsll)$ & $1.0  < \qsq < 6.0 \gev^2$ & $\checkmark$ & - \\
    $A_\text{FB}(\BXsll)$ & $14.3  < \qsq < 25 \gev^2$ & $\checkmark$ & - \\
    $A_\text{FB}(\BXsll)$ & $1.0  < \qsq < 3.5 \gev^2$ & - & \checkmark \\
    $A_\text{FB}(\BXsll)$ & $3.5  < \qsq < 6.0 \gev^2$ & - & \checkmark \\
    $A_\text{FB}(\BXsll)$ & $\qsq > 14.4 \gev^2$ & - & \checkmark \\
    $d\BF/d\qsq(\BtoKmumu)$ & $1.0  < \qsq < 6.0 \gev^2$ & - & \checkmark \\
    $d\BF/d\qsq(\BtoKmumu)$ & $1.1  < \qsq < 6.0 \gev^2$ & \checkmark & \checkmark \\
    $d\BF/d\qsq(\BtoKmumu)$ & $\qsq > 14.4 \gev^2$ & - & \checkmark \\
    $d\BF/d\qsq(\BtoKmumu)$ & $15.0  < \qsq < 22.0 \gev^2$ & \checkmark & \checkmark \\
    $d\BF/d\qsq(\BtoKstmumu)$ & $1.1  < \qsq < 6.0 \gev^2$ & \checkmark & \checkmark \\
    $d\BF/d\qsq(\BtoKstmumu)$ & $15.0  < \qsq < 19.0 \gev^2$ & \checkmark & \checkmark \\
    $d\BF/d\qsq(\BtoKstmumu)$ & $ \qsq > 14.4 \gev^2$ & - & \checkmark \\
    $d\BF/d\qsq(\Bstophimumu)$ & $1.0  < \qsq < 6.0 \gev^2$ & \checkmark & \checkmark \\
    $d\BF/d\qsq(\Bstophimumu)$ & $15.0  < \qsq < 19.0 \gev^2$ & \checkmark & \checkmark \\
    $F_L, A_\text{FB}, S_{4,5}(\BtoKstmumu)$ & $1.1  < \qsq < 2.5 \gev^2$ & \checkmark & \checkmark \\
    $F_L, A_\text{FB}, S_{4,5}(\BtoKstmumu)$ & $2.5  < \qsq < 4.0 \gev^2$ & \checkmark & \checkmark \\
    $F_L, A_\text{FB}, S_{4,5}(\BtoKstmumu)$ & $4.0  < \qsq < 6.0 \gev^2$ & \checkmark & \checkmark \\
    $F_L, A_\text{FB}, S_{4,5}(\BtoKstmumu)$ & $15  < \qsq < 19.0 \gev^2$ & \checkmark & \checkmark \\
    $P^{\prime}_{4,5}(\BtoKstmumu)$ & $0.1  < \qsq < 4.0 \gev^2$ & \checkmark & - \\
    $P^{\prime}_{4,5}(\BtoKstmumu)$ & $1.0  < \qsq < 2.5 \gev^2$ & - & \checkmark \\
    $P^{\prime}_{4,5}(\BtoKstmumu)$ & $2.5  < \qsq < 4.0 \gev^2$ & - & \checkmark \\
    $P^{\prime}_{4,5}(\BtoKstmumu)$ & $4.0  < \qsq < 6.0 \gev^2$ & - & \checkmark \\
    $P^{\prime}_{4,5}(\BtoKstmumu)$ & $14.18  < \qsq < 19.0 \gev^2$ & \checkmark & \checkmark \\
                \hline
  \end{tabular}\end{center}
\label{tab:inputsSL}
\end{table*}

\begin{table*}
\caption{Specific inputs for the scans of the semi-leptonic Wilson coefficients. It is indicated if the observable is included in the current average and/or the extrapolations to future milestones.}
\begin{center}\begin{tabular}{lccc}
    \hline
    Observable & $\qsq$ interval & current average & extrapolations \\
               \hline
        \multicolumn{4}{c}{Scans including primed coefficients} \\
    \hline
    $S_{3}(\BtoKstmumu)$ & $1.1  < \qsq < 2.5 \gev^2$ & \checkmark & \checkmark \\
    $S_{3}(\BtoKstmumu)$ & $2.5  < \qsq < 4.0 \gev^2$ & \checkmark & \checkmark \\
    $S_{3}(\BtoKstmumu)$ & $4.0  < \qsq < 6.0 \gev^2$ & \checkmark & \checkmark \\
    $S_{3}(\BtoKstmumu)$ & $15  < \qsq < 19.0 \gev^2$ & \checkmark & \checkmark \\
               \hline
        \multicolumn{4}{c}{Scans including \Cten} \\
    \hline
        $\BF(B_s\to\mu\mu)$ &  & \checkmark & \checkmark \\
       \hline
        \multicolumn{4}{c}{Scans including electron information} \\
    \hline
    $P^{\prime}_{4,5}(\BtoKstee)$ & $0.1  < \qsq < 4.0 \gev^2$ & \checkmark & - \\
    $P^{\prime}_{4,5}(\BtoKstee)$ & $1.0  < \qsq < 2.5 \gev^2$ & - & \checkmark \\
    $P^{\prime}_{4,5}(\BtoKstee)$ & $2.5  < \qsq < 4.0 \gev^2$ & - & \checkmark \\
    $P^{\prime}_{4,5}(\BtoKstee)$ & $4.0  < \qsq < 6.0 \gev^2$ & - & \checkmark \\
    $P^{\prime}_{4,5}(\BtoKstee)$ & $14.18  < \qsq < 19.0 \gev^2$ & \checkmark & \checkmark \\
    $R(X_s)$ & $1.0  < \qsq < 6.0 \gev^2$ & \checkmark & \checkmark \\
    $R(X_s)$ & $\qsq > 14.4 \gev^2$ & \checkmark & \checkmark \\
    $\Rk$ & $1.0  < \qsq < 6.0 \gev^2$ & $\checkmark$ & \checkmark \\
    $\Rk$ & $\qsq > 14.4 \gev^2$ & - & \checkmark \\
    $\Rk$ & $15.0  < \qsq < 22.0 \gev^2$ & - & \checkmark \\
    $\Rkst$ & $0.045  < \qsq < 1.1 \gev^2$ & \checkmark & \checkmark \\
    $\Rkst$ & $1.1  < \qsq < 6.0 \gev^2$ & \checkmark & \checkmark \\
    $\Rkst$ & $15.0  < \qsq < 19.0 \gev^2$ & - & \checkmark \\
    $\Rkst$ & $ \qsq > 14.4 \gev^2$ & - & \checkmark \\
    $R(\phi)$ & $1.0  < \qsq < 6.0 \gev^2$ & - & \checkmark \\
    $R(\phi)$ & $15.0  < \qsq < 19.0 \gev^2$ & - & \checkmark \\
                \hline
  \end{tabular}\end{center}
\label{tab:inputsSLspecifics}
\end{table*}

\begin{table*}
\caption{Inputs for the scans of the electromagnetic dipole Wilson coefficients. It is indicated if the observable is included in the current average and/or the extrapolations to future milestones.}
\begin{center}\begin{tabular}{lccc}
    \hline
    Observable & $\qsq$ interval & current average & extrapolations \\
    \hline
    $\BF(\Bstophigamma)$ &  & \checkmark & \checkmark \\
    $\BF(\BtoKstpgamma)$ &  & \checkmark & \checkmark \\
    $\ACP(\BtoKstpgamma)$ &  & \checkmark & \checkmark \\
    $\BF(\BtoKstzgamma)$ &  & \checkmark & \checkmark \\
    $\ACP(\BtoKstzgamma)$ &  & \checkmark & \checkmark \\
    $\BF(\BtoXsgamma)$ &  & \checkmark & \checkmark \\
    $A^{\Delta\Gamma}(\Bstophigamma)$ &  & \checkmark & \checkmark \\
    $S_{K^*\gamma}$ &  & \checkmark & \checkmark \\
    $P_1(\BtoKstee)$ & $0.002  < \qsq < 1.12 \gev^2$ & \checkmark & \checkmark \\
    $A^{\rm Im}_T(\BtoKstee)$ & $0.002  < \qsq < 1.12 \gev^2$ & \checkmark & \checkmark \\
    $A_{7,8,9}(\BtoKstmumu)$ & $1.1  < \qsq < 2.5 \gev^2$ & \checkmark & \checkmark \\
    $A_{7,8,9}(\BtoKstmumu)$ & $2.5  < \qsq < 4.0 \gev^2$ & \checkmark & \checkmark \\
    $A_{7,8,9}(\BtoKstmumu)$ & $4.0  < \qsq < 6.0 \gev^2$ & \checkmark & \checkmark \\
    $A_{8,9}(\BtoKstmumu)$ & $15  < \qsq < 19.0 \gev^2$ & \checkmark & \checkmark \\
               \hline
  \end{tabular}\end{center}
\label{tab:inputsRad}
\end{table*}

\begin{table*}[h]
\caption{Summary of inclusive inputs for the current measurement and extrapolations of the Belle (II) inclusive measurements. For the published measurements, the appropriate reference is given. The extrapolated uncertainties comprise statistical and systematic uncertainties.}
\begin{center}
  \begin{threeparttable}
  \begin{tabular}{lcccc}
    \hline
    Observable & $\qsq$ interval & Measurement &  \multicolumn{2}{c}{Extrapolations}\\
     & & $0.7 \invab$ & $5 \invab$ & $50 \invab$ \\
   \hline
    $\BF(\BtoXsgamma)$ & &  $(3.06 \pm 0.11 \pm 0.24 \pm 0.09)\times10^{-4}$~\cite{Belle:2016ufb} & $3.9\%$ & $3.2\%$ \\
    $\BF(\BXsmumu)$ & $1.0  < \qsq < 3.5 \gev^2$  &  -  & $17\%$ & $7.4\%$  \\
    $\BF(\BXsmumu)$ & $3.5  < \qsq < 6.0 \gev^2$  &  -  & $14\%$ & $6.8\%$ \\
    $\BF(\BXsmumu)$ & $1.0  < \qsq < 6.0 \gev^2$  & $(0.66^{+ 0.82}_{- 0.76} \,^{+0.30}_{-0.24} \pm 0.07)\times 10^{-6}$~\cite{Lees:2013nxa} & - & - \\
    $\BF(\BXsmumu)$ & $\qsq  > 14.4 \gev^2$  & $(0.60^{+ 0.31}_{- 0.29} \,^{+0.05}_{-0.04})\times 10^{-6}$~\cite{Lees:2013nxa} & $12\%$ & $5.1\%$\\
    $A_\text{FB}(\BXsll)$ & $1.0  < \qsq < 6.0 \gev^2$  & $0.30 \pm 0.24 \pm 0.04$~\cite{Sato:2014pjr} &  -  &  -  \\
    $A_\text{FB}(\BXsll)$ & $14.3  < \qsq < 25.0 \gev^2$  & $0.28 \pm 0.15 \pm 0.02$~\cite{Sato:2014pjr} &  -  &  - \\
    $A_\text{FB}(\BXsll)$ & $1.0  < \qsq < 3.5 \gev^2$  &  -  & $15\%$ & $4.7\%$  \\
    $A_\text{FB}(\BXsll)$ & $3.5  < \qsq < 6.0 \gev^2$  &  -  & $12\%$ & $3.8\%$ \\
    $A_\text{FB}(\BXsll)$ & $\qsq > 14.4 \gev^2$  &  -  & $9.5\%$ & $3.1\%$ \\
   $R(X_s)$ & $1.0  < \qsq < 6.0 \gev^2$ & $0.34 \pm 0.43$~\cite{Lees:2013nxa}\tnote{\hyperlink{fnsix}{vi}} & $12\%$ & $4\%$\\
   $R(X_s)$ & $\qsq > 14.4 \gev^2$ & $1.07 \pm 0.64$~\cite{Lees:2013nxa}\tnote{\hyperlink{fnsix}{vi}} & $17\%$ & $5.3\%$\\
             \hline
  \end{tabular}
  \begin{tablenotes}
 \hypertarget{fnsix}{\item[vi]{Calculated from Table I assuming fully correlated model and systematic uncertainties.}}
   \end{tablenotes}
  \end{threeparttable}
\end{center}
\label{tab:extrapolationsBelleincl}
\end{table*}

\begin{table*}[h]
\caption{Summary of exclusive inputs for the current measurement and extrapolations of the Belle (II) exclusive measurements. For the published measurements, the appropriate reference is given. The extrapolated uncertainties comprise statistical and systematic uncertainties.}
\begin{center}\begin{tabular}{lccccc}
    \hline
    Observable & $\qsq$ interval & Measurement &  \multicolumn{2}{c}{Extrapolations}\\
     & & $0.7 \invab$ & $5 \invab$ & $50 \invab$ \\
       \hline
    $\BF(\Bstophigamma)$ &  & $(3.6 \pm 0.5 \pm 0.3 \pm 0.6 )\times 10^{-5}$~\cite{Dutta:2014sxo} & -  &  -  \\
    $\BF(\BtoKstpgamma)$ & & $(39.2^{+1.3}_{-1.2})\times 10^{-6}$~\cite{HFAG_Kstg} & $1.8\times10^{-6}$ & $1.8\times10^{-6}$ \\
    $\ACP(\BtoKstpgamma)$ & & $0.012 \pm 0.023$~\cite{HFAG_ACP} & $0.0081$ & $0.0029$ \\
    $\BF(\BtoKstzgamma)$ & &  $(41.8 \pm 1.2)\times 10^{-6}$~\cite{HFAG_Kstg} & $1.5\times10^{-6}$ & $1.5\times10^{-6}$\\
    $\ACP(\BtoKstzgamma)$ & & $-0.007 \pm 0.011$~\cite{HFAG_ACP} & $0.0058$ & $0.0021$  \\
    $\SKstgamma$ &  & $-0.16 \pm 0.22$~\cite{HFAG_SCP} & $0.09$ & $0.03$ \\
    $\ATtwo(\BtoKstee)$ & $0.002  < \qsq < 1.12 \gev^2$ &  -  & $0.21$ & $0.066$ \\
    $\ATIm(\BtoKstee)$ & $0.002  < \qsq < 1.12 \gev^2$ &  -  & 0.20  & 0.064\\
    $\Rk$ & $1.0  < \qsq < 6.0 \gev^2$ &  -  & $11\%$ & $3.6\%$  \\
    $\Rk$ & $\qsq > 14.4 \gev^2$  &  -   & $12\%$ & $3.6\%$ \\
    $\Rkst$ & $1.1  < \qsq < 6.0 \gev^2$  &  -   & $10\%$  & $3.2\%$  \\
    $\Rkst$ & $ \qsq > 14.4 \gev^2$ &  -   & $9.2\%$ & $2.8\%$ \\
     $d\BF/d\qsq(\BtoKmumu)$ & $1.0  < \qsq < 6.0 \gev^2$   &  -   & $10\%$  & $4\%$ \\
    $d\BF/d\qsq(\BtoKmumu)$ & $\qsq > 14.4 \gev^2$   &  -   & $10\%$ & $4\%$ \\
    $d\BF/d\qsq(\BtoKstmumu)$ & $1.1  < \qsq < 6.0 \gev^2$  &  -   & $10\%$ & $4\%$ \\
    $d\BF/d\qsq(\BtoKstmumu)$ & $ \qsq > 14.4 \gev^2$  &  -   & $10\%$ & $4\%$ \\
               \hline
  \end{tabular}\end{center}
\label{tab:extrapolationsBelleexcl1}
\end{table*}

\begin{table*}[h]
\caption{Summary of exclusive inputs for the current measurement and extrapolations of the Belle (II) exclusive measurements. For the published measurements, the appropriate reference is given. The extrapolated uncertainties comprise statistical and systematic uncertainties.}
\begin{center}\begin{tabular}{lcccc}
    \hline
    Observable & $\qsq$ interval & Measurement &  \multicolumn{2}{c}{Extrapolations} \\
     & & $0.7 \invab$ & $5 \invab$ & $50 \invab$  \\
       \hline
    $P^{\prime}_{4}(\BtoKstmumu)$ & $0.1  < \qsq < 4.0 \gev^2$  & $-0.38^{+0.50}_{-0.48} \pm 0.12$~\cite{Wehle:2016yoi} & - & - \\
    $P^{\prime}_{4}(\BtoKstmumu)$ & $1.0  < \qsq < 2.5 \gev^2$  & - & $0.27$ & $0.08$  \\
    $P^{\prime}_{4}(\BtoKstmumu)$ & $2.5  < \qsq < 4.0 \gev^2$  & - & $0.24$ & $0.08$ \\
    $P^{\prime}_{4}(\BtoKstmumu)$ & $4.0  < \qsq < 6.0 \gev^2$  & - & $0.19$ & $0.06$ \\
    $P^{\prime}_{4}(\BtoKstmumu)$ & $14.18  < \qsq < 19.0 \gev^2$  & $-0.10^{+0.39}_{-0.39} \pm 0.07$~\cite{Wehle:2016yoi} & $0.13$& $0.04$\\
    $P^{\prime}_{5}(\BtoKstmumu)$ & $0.1  < \qsq < 4.0 \gev^2$  & $\phantom{-}0.42^{+ 0.39}_{-0.39} \pm 0.14$~\cite{Wehle:2016yoi} & - & -\\
    $P^{\prime}_{5}(\BtoKstmumu)$ & $1.0  < \qsq < 2.5 \gev^2$  & - & $0.25$ & $0.08$  \\
    $P^{\prime}_{5}(\BtoKstmumu)$ & $2.5  < \qsq < 4.0 \gev^2$  & - &$0.23$ & $0.07$ \\
    $P^{\prime}_{5}(\BtoKstmumu)$ & $4.0  < \qsq < 6.0 \gev^2$  & - &$0.18$ & $0.06$ \\
    $P^{\prime}_{5}(\BtoKstmumu)$ & $14.18  < \qsq < 19.0 \gev^2$  & $-0.13^{+0.39}_{-0.35} \pm 0.06$~\cite{Wehle:2016yoi} & $0.11$ &$0.04$\\
    $P^{\prime}_{4}(\BtoKstee)$ & $0.1  < \qsq < 4.0 \gev^2$  & $\phantom{-}0.34^{+0.41}_{-0.45} \pm 0.11$~\cite{Wehle:2016yoi} & - & - \\
    $P^{\prime}_{4}(\BtoKstee)$ & $1.0  < \qsq < 2.5 \gev^2$  & - &$0.24$ & $0.07$ \\
    $P^{\prime}_{4}(\BtoKstee)$ & $2.5  < \qsq < 4.0 \gev^2$  & - &$0.22$ & $0.07$  \\
    $P^{\prime}_{4}(\BtoKstee)$ & $4.0  < \qsq < 6.0 \gev^2$  & - &$0.18$ & $0.06$ \\
    $P^{\prime}_{4}(\BtoKstee)$ & $14.18  < \qsq < 19.0 \gev^2$  & $-0.15^{+0.41}_{-0.40} \pm 0.04$~\cite{Wehle:2016yoi} & $0.16$ & $0.05$\\
    $P^{\prime}_{5}(\BtoKstee)$ & $0.1  < \qsq < 4.0 \gev^2$  & $\phantom{-}0.51^{+0.39}_{-0.46} \pm 0.09$~\cite{Wehle:2016yoi} & - & -  \\
    $P^{\prime}_{5}(\BtoKstee)$ & $1.0  < \qsq < 2.5 \gev^2$  & - &$0.23$ & $0.07$   \\
    $P^{\prime}_{5}(\BtoKstee)$ & $2.5  < \qsq < 4.0 \gev^2$  & - &$0.21$ & $0.07$  \\
    $P^{\prime}_{5}(\BtoKstee)$ & $4.0  < \qsq < 6.0 \gev^2$  & - &$0.17$ & $0.06$  \\
    $P^{\prime}_{5}(\BtoKstee)$ & $14.18  < \qsq < 19.0 \gev^2$  & $-0.91^{+ 0.36}_{-0.30} \pm 0.03$~\cite{Wehle:2016yoi} & $0.14$ & $0.04$ \\
               \hline
  \end{tabular}\end{center}
\label{tab:extrapolationsBelleexcl2}
\end{table*}

\begin{sidewaystable*}[h]
\caption{Summary of inputs for the current measurement and extrapolations of the \lhcb measurements. For the published measurements, the appropriate reference is given. The extrapolated uncertainties comprise solely statistical uncertainties.}
\begin{center}
  \begin{threeparttable}
  \begin{tabular}{lccccc}
    \hline
    Observable & $\qsq$ interval & Measurement &  \multicolumn{3}{c}{Extrapolations} \\
     & & $3 \invfb$ & $8 \invfb$ & $22 \invfb$ & $50 \invfb$ \\
    \hline
    $\BF(\Bsmumu)$ &  & $(2.9^{+1.1}_{-1.0} \,^{+0.3}_{-0.1})\times 10^{-9}$~\cite{Aaij:2013aka}\tnote{\hyperlink{fnseven}{vii}} &$0.44\times 10^{-9}$ & $0.24\times 10^{-9}$ & $0.16\times 10^{-9}$  \\
    $\BF(\Bstophigamma)$ &  & $(3.5 \pm 0.3)\times 10^{-5}$~\cite{Aaij:2012ita}\tnote{\hyperlink{fneight}{viii}} &$0.3\times 10^{-5}$ & $0.3\times 10^{-5}$ & $0.3\times 10^{-5}$ \\
    $\ADeltaG(\Bstophigamma)$ &  & $-0.98^{+0.46}_{-0.52} \,^{+0.23}_{-0.20}$~\cite{Aaij:2016ofv} & 0.27 & 0.11 & 0.07  \\
    $\ATtwo(\BtoKstee)$ & $0.002  < \qsq < 1.12 \gev^2$ & $-0.23 \pm 0.23 \pm 0.05$~\cite{Aaij:2015dea} & $0.120$ & $0.053$ & 0.033 \\
    $\ATIm(\BtoKstee)$ & $0.002  < \qsq < 1.12 \gev^2$ & $0.14 \pm 0.22 \pm 0.05$~\cite{Aaij:2015dea} & 0.115  & 0.050 & 0.032   \\
    $R(\phi)$ & $1.0  < \qsq < 6.0 \gev^2$ &  -   & 0.159 & 0.086 & 0.056 \\
    $R(\phi)$ & $15.0  < \qsq < 19.0 \gev^2$ &  -  & 0.137 & 0.074 & 0.048 \\
    $\Rk$ & $1.0  < \qsq < 6.0 \gev^2$ & $0.745^{+0.090}_{-0.074} \pm 0.036$~\cite{Aaij:2014ora} & $0.046$ & $0.025$ & 0.016  \\
    $\Rk$ & $15.0  < \qsq < 22.0 \gev^2$ &  -  & 0.043 & 0.023 & 0.015  \\
    $\Rkst$ & $0.045  < \qsq < 1.1 \gev^2$  & $0.66^{+0.11}_{-0.07} \pm 0.03$~\cite{Aaij:2017vbb} & 0.048 & 0.026 & 0.017   \\
    $\Rkst$ & $1.1  < \qsq < 6.0 \gev^2$  & $0.69^{+0.11}_{-0.07} \pm 0.05$~\cite{Aaij:2017vbb} & 0.053 & 0.028 & 0.019   \\
    $\Rkst$ & $15.0  < \qsq < 19.0 \gev^2$  &  -   & 0.061 & 0.033 & 0.021 \\
    $d\BF/d\qsq(\BtoKmumu)$ & $1.1  < \qsq < 6.0 \gev^2$  & $(24.2 \pm 0.7 \pm 1.2) \times 10^{-9}$~\cite{Aaij:2014pli}  & $0.7\times 10^{-9}$ &  $0.4\times 10^{-9}$ & $0.3\times 10^{-9}$ \\
    $d\BF/d\qsq(\BtoKmumu)$ & $15.0  < \qsq < 22.0 \gev^2$   & $(12.1 \pm 0.4 \pm 0.6)\times 10^{-9}$~\cite{Aaij:2014pli} & $0.4\times 10^{-9}$ &  $0.2\times 10^{-9}$ & $0.1\times 10^{-9}$  \\
    $d\BF/d\qsq(\BtoKstmumu)$ & $1.1  < \qsq < 6.0 \gev^2$  & $(0.342^{+0.017}_{-0.017} \pm  0.009  \pm 0.023)\times 10^{-7}$~\cite{Aaij:2016flj} & $0.015\times 10^{-7}$ &  $0.008\times 10^{-7}$ & $0.005\times 10^{-7}$ \\
    $d\BF/d\qsq(\BtoKstmumu)$ & $15.0  < \qsq < 19.0 \gev^2$  & $(0.436^{+0.018}_{-0.019} \pm  0.007 \pm 0.030)\times 10^{-7}$~\cite{Aaij:2016flj} & $0.018\times 10^{-7}$ & $0.010\times 10^{-7}$  & $0.006\times 10^{-7}$ \\
     $d\BF/d\qsq(\Bstophimumu)$ & $1.0  < \qsq < 6.0 \gev^2$  &  $(2.58^{+0.33}_{-0.31} \pm  0.08 \pm 0.19)\times 10^{-8}$~\cite{Aaij:2015esa}  & $0.20\times 10^{-8}$ & $0.11\times 10^{-8}$ & $0.07\times 10^{-8}$\\
     $d\BF/d\qsq(\Bstophimumu)$ & $15.0  < \qsq < 19.0 \gev^2$  &  $(4.04^{+0.39}_{-0.38} \pm  0.13 \pm 0.30)\times 10^{-8}$~\cite{Aaij:2015esa}  & $0.26\times 10^{-8}$ & $0.14\times 10^{-8}$ & $0.09\times 10^{-8}$ \\
            \hline
  \end{tabular}
  \begin{tablenotes}
 \hypertarget{fnseven}{\item[vii]{The extrapolations of the \lhcb measurement of \Bsmumu do not rely on the quoted measured branching fraction on the $3\invfb$ dataset of Run 1~\cite{Aaij:2013aka} but on the update including partial data from Run 2 on $4.4\invfb$~\cite{Aaij:2017vad}.}}
\hypertarget{fneight}{\item[viii]{This measurement has been performed on 1 \invfb and has been extrapolated to 3\invfb under the assumption that the uncertainty arising from $f_s/f_d$ is irreducible}}
  \end{tablenotes}
  \end{threeparttable}
\end{center}
\label{tab:extrapolationsLHCb1}
\end{sidewaystable*}

\begin{table*}[h]
\caption{Summary of inputs for the current measurement and extrapolations of the \lhcb measurements. For the published measurements, the appropriate reference is given. The extrapolated uncertainties comprise solely statistical uncertainties.}
\begin{center}\begin{tabular}{lccccc}
    \hline
    Observable & $\qsq$ interval & Measurement &  \multicolumn{3}{c}{Extrapolations}  \\
     & & $3 \invfb$ & $8 \invfb$ & $22 \invfb$ & $50 \invfb$   \\
    \hline
    $S_{3}(\BtoKstmumu)$ & $1.1  < \qsq < 2.5 \gev^2$  & $-0.077^{+0.087}_{-0.105} \pm 0.005$~\cite{Aaij:2015oid} & 0.049 & 0.027 & 0.017 \\
    $S_{4}(\BtoKstmumu)$ & $1.1  < \qsq < 2.5 \gev^2$  & $-0.077^{+0.111}_{-0.113}  \pm 0.005$~\cite{Aaij:2015oid} & 0.057 & 0.031 & 0.020  \\
    $S_{5}(\BtoKstmumu)$ & $1.1  < \qsq < 2.5 \gev^2$  & $\phantom{-}0.137^{+0.099 }_{-0.094}  \pm 0.009$~\cite{Aaij:2015oid} & 0.050 & 0.027 & 0.018 \\
    $F_L(\BtoKstmumu)$ & $1.1  < \qsq < 2.5 \gev^2$  & $\phantom{-}0.660^{+0.083}_{- 0.077}  \pm 0.022$~\cite{Aaij:2015oid}  &  0.042& 0.023 & 0.015    \\
    $A_\text{FB}(\BtoKstmumu)$ & $1.1  < \qsq < 2.5 \gev^2$  & $-0.191^{+0.068}_{ -0.080}  \pm 0.012$~\cite{Aaij:2015oid} & 0.038 & 0.021 & 0.014  \\
    $A_7(\BtoKstmumu)$ & $1.1  < \qsq < 2.5 \gev^2$  & $-0.087^{+0.091}_{ -0.093}  \pm 0.004$~\cite{Aaij:2015oid} & 0.047 & 0.025 & 0.017   \\
    $A_8(\BtoKstmumu)$ & $1.1  < \qsq < 2.5 \gev^2$  & $-0.044^{+0.108}_{ -0.113}  \pm 0.005$~\cite{Aaij:2015oid} & 0.057  & 0.031 & 0.020  \\
    $A_9(\BtoKstmumu)$ & $1.1  < \qsq < 2.5 \gev^2$  & $-0.004^{+0.092}_{ -0.098}  \pm 0.005$~\cite{Aaij:2015oid} & 0.049 & 0.026 & 0.017  \\
    $S_{3}(\BtoKstmumu)$ & $2.5  < \qsq < 4.0 \gev^2$  & $\phantom{-}0.035^{+0.098}_{-0.089} \pm 0.007$~\cite{Aaij:2015oid} &0.048  & 0.026 & 0.017   \\
    $S_{4}(\BtoKstmumu)$ & $2.5  < \qsq < 4.0 \gev^2$  & $-0.234^{+0.127}_{-0.144} \pm 0.006$~\cite{Aaij:2015oid} & 0.070  & 0.038 & 0.025    \\
    $S_{5}(\BtoKstmumu)$ & $2.5  < \qsq < 4.0 \gev^2$  & $-0.022^{+0.110}_{-0.103}  \pm 0.008$~\cite{Aaij:2015oid} & 0.055 & 0.030 & 0.019   \\
    $F_L(\BtoKstmumu)$ & $2.5  < \qsq < 4.0 \gev^2$   & $\phantom{-}0.876^{+0.109}_{-0.097}  \pm 0.017$~\cite{Aaij:2015oid} & 0.053  & 0.029 & 0.019  \\
    $A_\text{FB}(\BtoKstmumu)$ & $2.5  < \qsq < 4.0 \gev^2$   & $-0.118^{+0.082}_{-0.090}  \pm 0.007$~\cite{Aaij:2015oid} & 0.044 & 0.024 & 0.016   \\
    $A_7(\BtoKstmumu)$ & $2.5  < \qsq < 4.0 \gev^2$  & $-0.032^{+0.109}_{ -0.115}  \pm 0.005$~\cite{Aaij:2015oid} & 0.057 & 0.031 & 0.020   \\
    $A_8(\BtoKstmumu)$ & $2.5  < \qsq < 4.0 \gev^2$  & $-0.071^{+0.124}_{ -0.131}  \pm 0.006$~\cite{Aaij:2015oid} & 0.065 & 0.035 & 0.023   \\
    $A_9(\BtoKstmumu)$ & $2.5  < \qsq < 4.0 \gev^2$  & $-0.228^{+0.114}_{ -0.152}  \pm 0.007$~\cite{Aaij:2015oid} & 0.068 & 0.037 & 0.024  \\
    $S_{3}(\BtoKstmumu)$ & $4.0  < \qsq < 6.0 \gev^2$   & $\phantom{-}0.035^{+0.069}_{-0.068}  \pm 0.007$~\cite{Aaij:2015oid} & 0.035 &  0.019 & 0.012    \\
    $S_{4}(\BtoKstmumu)$ & $4.0  < \qsq < 6.0 \gev^2$   & $-0.219^{+0.086}_{-0.084}  \pm 0.008$~\cite{Aaij:2015oid} & 0.044  & 0.024 & 0.015    \\
    $S_{5}(\BtoKstmumu)$ & $4.0  < \qsq < 6.0 \gev^2$   & $-0.146^{+0.077}_{-0.078}  \pm 0.011$~\cite{Aaij:2015oid} & 0.040 & 0.022 & 0.014   \\
    $F_L(\BtoKstmumu)$ & $4.0  < \qsq < 6.0 \gev^2$   & $\phantom{-}0.611^{+0.052}_{-0.053}  \pm 0.017$~\cite{Aaij:2015oid} & 0.028 & 0.015 & 0.010   \\
    $A_\text{FB}(\BtoKstmumu)$ & $4.0  < \qsq < 6.0 \gev^2$   & $\phantom{-}0.025^{+0.051}_{-0.052}  \pm 0.004$~\cite{Aaij:2015oid}  & 0.027 & 0.014 & 0.009  \\
    $A_7(\BtoKstmumu)$ & $4.0  < \qsq < 6.0 \gev^2$  & $\phantom{-}0.041^{+0.083}_{ -0.082}  \pm 0.004$~\cite{Aaij:2015oid} & 0.042 & 0.023 & 0.015  \\
    $A_8(\BtoKstmumu)$ & $4.0  < \qsq < 6.0 \gev^2$  & $\phantom{-}0.004^{+0.093}_{ -0.095}  \pm 0.005$~\cite{Aaij:2015oid} & 0.048 & 0.026 & 0.017   \\
    $A_9(\BtoKstmumu)$ & $4.0  < \qsq < 6.0 \gev^2$  & $\phantom{-}0.062^{+0.078}_{ -0.072}  \pm 0.004$~\cite{Aaij:2015oid} & 0.038 & 0.021  & 0.014   \\
    $S_{3}(\BtoKstmumu)$ & $15  < \qsq < 19.0 \gev^2$  & $-0.163^{+0.033}_{-0.033}  \pm 0.009$~\cite{Aaij:2015oid} & 0.017 &0.009  & 0.006    \\
    $S_{4}(\BtoKstmumu)$ & $15  < \qsq < 19.0 \gev^2$  & $-0.284^{+0.038}_{-0.041}  \pm 0.007$~\cite{Aaij:2015oid} & 0.021 & 0.011 & 0.007    \\
    $S_{5}(\BtoKstmumu)$ & $15  < \qsq < 19.0 \gev^2$  & $-0.325^{+0.036}_{-0.037}  \pm 0.009$~\cite{Aaij:2015oid} & 0.019 & 0.011 & 0.007   \\
    $F_L(\BtoKstmumu)$ & $15.0  < \qsq < 19.0 \gev^2$  & $\phantom{-}0.344^{+0.028}_{-0.030}  \pm 0.008$~\cite{Aaij:2015oid} & 0.015 & 0.008 & 0.005   \\
    $A_\text{FB}(\BtoKstmumu)$ & $15.0  < \qsq < 19.0 \gev^2$  & $\phantom{-}0.355^{+0.027}_{-0.027}  \pm 0.009$~\cite{Aaij:2015oid} & 0.015 & 0.008 & 0.005 \\
    $A_8(\BtoKstmumu)$ & $15.0  < \qsq < 19.0 \gev^2$  & $\phantom{-}0.025^{+0.048}_{ -0.047}  \pm 0.003$~\cite{Aaij:2015oid} & 0.025 & 0.013  & 0.009  \\
    $A_9(\BtoKstmumu)$ & $15.0  < \qsq < 19.0 \gev^2$  & $\phantom{-}0.061^{+0.043}_{ -0.044}  \pm 0.002$~\cite{Aaij:2015oid} & 0.023 & 0.012 & 0.008 \\
            \hline
  \end{tabular}\end{center}
\label{tab:extrapolationsLHCb2}
\end{table*}

\begin{sidewaystable*}[h]
\caption{Summary of the Belle (II) inclusive measurements and the corresponding central values for the SM and the different NP scenarios as obtained with the \texttt{flavio} package.}
\begin{center}
\resizebox{\linewidth}{!}{
\begin{tabular}{lccccccc}
    \hline
   Observable & $\qsq$ interval & SM &  \multicolumn{1}{c}{(\Cnine, \Cten)} &  \multicolumn{1}{c}{(\Cninep, \Ctenp)} & \multicolumn{1}{c}{(\Cnine, \Cnineee)} &  \multicolumn{1}{c}{(\ReCsevp, \ImCsevp)} &  \multicolumn{1}{c}{(\ReCsev, \ImCsev)}\\
    & & & $(-0.8, 0.6)$ & $(0.8, 0.2)$ & $ (-0.8, 0.4)$ & $(0.02, -0.06)$ & $(-0.050, -0.075)$\\
   \hline
    $\BF(\BtoXsgamma) \, [10^{-4}]$ & & $3.30$  &  -  &  -  & & $3.36$ & $4.14$\\
    $\BF(\BXsmumu) \, [10^{-6}]$ & $1.0  < \qsq < 3.5 \gev^2$ & $0.92$ & $0.67$ & $0.94$ & $0.81$ & - &  - \\
    $\BF(\BXsmumu)  \, [10^{-6}]$ & $3.5  < \qsq < 6.0 \gev^2$ & $0.76$ & $0.52$ & $0.77$ & $0.64$ &  -  & -  \\
    $\BF(\BXsmumu)  \, [10^{-6}]$ & $\qsq  > 14.4 \gev^2$ & $0.31$ & $0.22$ & $0.32$ & $0.27$ &  -  &  - \\
    $A_\text{FB}(\BXsll)$ & $1.0  < \qsq < 3.5 \gev^2$  & $-0.080$ & $-0.098$ & $-0.079$ & $-0.088$ &  -  & -  \\
    $A_\text{FB}(\BXsll)$ & $3.5  < \qsq < 6.0 \gev^2$  & $0.065$& $0.047$ & $0.066$ & $0.055$ &  -  & -  \\
    $A_\text{FB}(\BXsll)$ & $\qsq > 14.4 \gev^2$ & $-0.066$ & $-0.061$ & $-0.063$ & $-0.062$ &  -  & -  \\
    $R(X_s)$ & $1.0  < \qsq < 6.0 \gev^2$ & $0.963$ &  -  &  -  & $0.772$ &  -  &  - \\
    $R(X_s)$ & $\qsq > 14.4 \gev^2$ & $1.149$ & - & - & $0.923$ & - & -\\
            \hline
  \end{tabular}}
  \end{center}
\label{tab:NPBelleincl}
\end{sidewaystable*}

\begin{sidewaystable*}[h]
\caption{Summary of the Belle (II) exclusive measurements and the corresponding central values for the SM and the different NP scenarios as obtained with the \texttt{flavio} package.}
\begin{center}
\resizebox{\linewidth}{!}{
\begin{tabular}{lccccccc}
    \hline
        Observable & $\qsq$ interval & SM &  \multicolumn{1}{c}{(\Cnine, \Cten)} &  \multicolumn{1}{c}{(\Cninep, \Ctenp)} & \multicolumn{1}{c}{(\Cnine, \Cnineee)} &  \multicolumn{1}{c}{(\ReCsevp, \ImCsevp)} &  \multicolumn{1}{c}{(\ReCsev, \ImCsev)}\\
    &  & & $(-1.4, 0.4)$ & $(0.4, 0.2)$ & $ (-1.4, -0.7)$ & $(0.08, 0.00)$ & $(-0.050, 0.050)$\\       \hline
    $\BF(\BtoKstpgamma) \, [10^{-6}]$ & & $41.8$ &  -  &  -  &  -  & $43.5$ & $54.6$\\
    $\ACP(\BtoKstpgamma)$ &  & 0.0053 &  -  &  -  &  -  & $0.0051$ & $-0.0114$\\
    $\BF(\BtoKstzgamma) \, [10^{-6}]$ & & $42.1$ & -  &  -  &  -  & $43.6$ & $54.5$\\
    $\ACP(\BtoKstzgamma)$ &  & 0.0036 &  -  &  -  &  -  & $0.0034$ & $-0.0116$\\
    $\SKstgamma$ & & $-0.02$ & -   &  -  &  -  & $0.27$ & $-0.02$ \\
    $\ATtwo(\BtoKstee)$ & $0.002  < \qsq < 1.12 \gev^2$ & $0.037$ &  -  &  -  &  -  & $-0.379$ & $0.032$\\
    $\ATIm(\BtoKstee)$ & $0.002  < \qsq < 1.12 \gev^2$ & $0.000$ & -  &  -  &  -  & $-0.003$ & $0.004$\\
    $\Rk$ & $1.0  < \qsq < 6.0 \gev^2$ & $1.000$ & -  &  -  & $0.857$ &  -  &  -  \\
    $\Rk$ & $\qsq > 14.4 \gev^2$  & $1.003$ &  -  &  -  & $0.861$ &  -  &  -  \\
    $\Rkst$ & $1.1  < \qsq < 6.0 \gev^2$  & $0.996$  & -  &  -  & $0.908$ &  -  &  -   \\
    $\Rkst$ & $ \qsq > 14.4 \gev^2$ & $0.998$ & -  &  -  & $0.861$ &  -  &  -   \\
     $d\BF/d\qsq(\BtoKmumu) \, [10^{-9}]$ & $1.0  < \qsq < 6.0 \gev^2$ & $36.5$ & $23.0$ & $38.3$ & $26.5$ &  -  &  -  \\
    $d\BF/d\qsq(\BtoKmumu) \, [10^{-9}]$ & $\qsq > 14.4 \gev^2$ & $11.9$ & $7.5$ & $12.5$ & $8.7$ &  -  &  -  \\
    $d\BF/d\qsq(\BtoKstmumu)  \, [10^{-7}]$ & $1.1  < \qsq < 6.0 \gev^2$ & $0.475$ & $0.334$ & $0.467$ & $0.387$ &  -  &  -   \\
    $d\BF/d\qsq(\BtoKstmumu) \, [10^{-7}]$ & $ \qsq > 14.4 \gev^2$ & $0.272$ & $0.172$ & $0.265$ & $0.200$ &  -  &  -  \\
               \hline
  \end{tabular}}
  \end{center}
\label{tab:NPBelleexcl1}
\end{sidewaystable*}

\begin{sidewaystable*}[ht]
\caption{Summary of the Belle (II) exclusive measurements and the corresponding central values for the SM and the different NP scenarios as obtained with the \texttt{flavio} package.}
\begin{center}
\resizebox{\linewidth}{!}{
\begin{tabular}{lccccccc}
    \hline
        Observable & $\qsq$ interval & SM &  \multicolumn{1}{c}{(\Cnine, \Cten)} &  \multicolumn{1}{c}{(\Cninep, \Ctenp)} & \multicolumn{1}{c}{(\Cnine, \Cnineee)} &  \multicolumn{1}{c}{(\ReCsevp, \ImCsevp)} &  \multicolumn{1}{c}{(\ReCsev, \ImCsev)}\\
    & & & $(-1.4, 0.4)$ & $(0.4, 0.2)$ & $ (-1.4, -0.7)$ & $(0.08, 0.00)$ & $(-0.050, 0.050)$\\       \hline
    $P^{\prime}_{4}(\BtoKstmumu)$ & $1.0  < \qsq < 2.5 \gev^2$ & $-0.04$ & $-0.04$ & $-0.07$ & $-0.08$ &  -  &  - \\
    $P^{\prime}_{4}(\BtoKstmumu)$ & $2.5  < \qsq < 4.0 \gev^2$ & $-0.39$ & $-0.32$ & $-0.41$ & $-0.35$ &  -  &  - \\
    $P^{\prime}_{4}(\BtoKstmumu)$ & $4.0  < \qsq < 6.0 \gev^2$ & $-0.50$ & $-0.47$ & $-0.51$ & $-0.48$ &  -  &  - \\
    $P^{\prime}_{4}(\BtoKstmumu)$ & $14.18  < \qsq < 19.0 \gev^2$ & $-0.63$ & $-0.63$ & $-0.62$ & $-0.63$ &  -  &  - \\
    $P^{\prime}_{5}(\BtoKstmumu)$ & $1.0  < \qsq < 2.5 \gev^2$ & $-0.17$ & $0.49$ & $0.21$ & $0.48$ &  -  &  - \\
    $P^{\prime}_{5}(\BtoKstmumu)$ & $2.5  < \qsq < 4.0 \gev^2$ & $-0.50$ & $-0.04$ & $-0.48$ & $-0.04$ &  -  &  - \\
    $P^{\prime}_{5}(\BtoKstmumu)$ & $4.0  < \qsq < 6.0 \gev^2$ & $-0.76$ & $-0.43$ & $-0.75$ & $-0.40$ &  -  &  - \\
    $P^{\prime}_{5}(\BtoKstmumu)$ & $14.18  < \qsq < 19.0 \gev^2$ & $-0.62$ & $-0.55$ & $-0.64$ & $-0.52$ &  -  &  - \\
    $P^{\prime}_{4}(\BtoKstee)$ & $1.0  < \qsq < 2.5 \gev^2$   & $-0.04$ &  -  &  -  & $-0.05$ &  -  &  - \\
    $P^{\prime}_{4}(\BtoKstee)$ & $2.5  < \qsq < 4.0 \gev^2$ & $-0.39$  &  -  &  -  & $-0.37$ &  -  &  - \\
    $P^{\prime}_{4}(\BtoKstee)$ & $4.0  < \qsq < 6.0 \gev^2$ & $-0.50$  &  -  &  -  & $-0.49$ &  -  &  - \\
    $P^{\prime}_{4}(\BtoKstee)$ & $14.18  < \qsq < 19.0 \gev^2$ & $-0.63$ &  -  &  -  & $-0.63$ &  -  &  - \\
    $P^{\prime}_{5}(\BtoKstee)$ & $1.0  < \qsq < 2.5 \gev^2$ & $0.17$ &  -  &  -  & $0.32$ &  -  &  - \\
    $P^{\prime}_{5}(\BtoKstee)$ & $2.5  < \qsq < 4.0 \gev^2$  & $-0.50$ &  -  &  -  & $-0.28$ &  -  &  - \\
    $P^{\prime}_{5}(\BtoKstee)$ & $4.0  < \qsq < 6.0 \gev^2$  & $-0.75$ &  -  &  -  & $-0.61$ &  -  &  - \\
    $P^{\prime}_{5}(\BtoKstee)$ & $14.18  < \qsq < 19.0 \gev^2$ & $-0.62$ &  -  &  -  & $-0.59$ &  -  &  - \\
               \hline
  \end{tabular}}
  \end{center}
\label{tab:NPBelleexcl2}
\end{sidewaystable*}

\begin{sidewaystable*}[h]
\caption{Summary of the \lhcb measurements and the corresponding central values for the SM and the different NP scenarios as obtained with the \texttt{flavio} package.}
\begin{center}
\resizebox{\linewidth}{!}{
\begin{tabular}{lccccccc}
    \hline
            Observable & $\qsq$ interval & &  \multicolumn{1}{c}{(\Cnine, \Cten)} &  \multicolumn{1}{c}{(\Cninep, \Ctenp)} & \multicolumn{1}{c}{(\Cnine, \Cnineee)} &  \multicolumn{1}{c}{(\ReCsevp, \ImCsevp)} &  \multicolumn{1}{c}{(\ReCsev, \ImCsev)}\\
    & & & $(-1.0, 0.0)$ & $(-0.2, -0.2)$ & $ (-1.0, 0.0)$ & $(0.00, 0.04)$ & $(-0.075, 0.000)$\\
        \hline
    $\BF(\Bsmumu) \, [10^{-9}]$ & & $3.60$  &   $3.60$   &   $3.27$  &  -  &- & - \\
    $\BF(\Bstophigamma) \, [10^{-5}]$ & & $4.1$  &  -  &  -  &  -  & $4.1$ & $5.8$ \\
    $\ADeltaG(\Bstophigamma)$ & & $0.03$  &  -  &  -  &  -  & $0.03$ & $0.03$\\
    $\ATtwo(\BtoKstee)$ & $0.002  < \qsq < 1.12 \gev^2$ & $0.037$ &  -  &  -  &  -  & $0.038$ & $0.033$\\
    $\ATIm(\BtoKstee)$ & $0.002  < \qsq < 1.12 \gev^2$ & $0.000$ &  -  &  -  &  -  & $-0.213$ & $0.032$\\
    $R(\phi)$ & $1.0  < \qsq < 6.0 \gev^2$ & $0.997$ & -  &  -  & $0.840$ &  -  &  - \\
    $R(\phi)$ & $15.0  < \qsq < 19.0 \gev^2$ & $0.998$ &  -  &  -  & $0.795$ &  -  &  -  \\
    $\Rk$ & $1.0  < \qsq < 6.0 \gev^2$ & $1.000$ & -  &  -  & $0.793$ &  -  &  - \\
    $\Rk$ & $15.0  < \qsq < 22.0 \gev^2$ & $1.003$ &  -  &  -  & $0.797$ &  - &  -  \\
    $\Rkst$ & $0.045  < \qsq < 1.1 \gev^2$  & $0.945$ & -  &  -  & $0.911$ &  -  &  -  \\
    $\Rkst$ & $1.1  < \qsq < 6.0 \gev^2$  & $0.996$ & -  &  -  & $0.846$ &  -  &  -  \\
    $\Rkst$ & $15.0  < \qsq < 19.0 \gev^2$  & $0.998$ &  -  &  -  & $0.796$ &  -  &  -  \\
    $d\BF/d\qsq(\BtoKmumu)  \, [10^{-9}]$ & $1.1  < \qsq < 6.0 \gev^2$ & $36.5$ & $28.9$ & $36.7$ & $28.9$ &  -  &  - \\
    $d\BF/d\qsq(\BtoKmumu)  \, [10^{-9}]$ & $15.0  < \qsq < 22.0 \gev^2$ & $15.6$ & $12.4$ & $15.6$ & $12.4$ &  -  &  - \\
    $d\BF/d\qsq(\BtoKstmumu)  \, [10^{-7}]$ & $1.1  < \qsq < 6.0 \gev^2$ & $0.475$ & $0.407$ & $0.477$ & $0.407$ &  -  &  - \\
    $d\BF/d\qsq(\BtoKstmumu)  \, [10^{-7}]$ & $15.0  < \qsq < 19.0 \gev^2$  & $0.596$ & $0.475$ & $0.594$ & $0.475$ &  -  &  -  \\
    $d\BF/d\qsq(\Bstophimumu)  \, [10^{-8}]$ & $1.0  < \qsq < 6.0 \gev^2$ & $5.38$ & $4.54$ & $5.37$ & $4.54$ &  -  &  - \\
    $d\BF/d\qsq(\Bstophimumu)  \, [10^{-8}]$ & $15.0  < \qsq < 19.0 \gev^2$  & $5.56$ & $4.43$ & $5.55$ & $4.43$ &  -  &  -  \\
            \hline
  \end{tabular}}
  \end{center}
\label{tab:NPLHCb1}
\end{sidewaystable*}

\begin{sidewaystable*}[h]
\caption{Summary of the \lhcb measurements and the corresponding central values for the SM and the different NP scenarios as obtained with the \texttt{flavio} package.}
\begin{center}
\resizebox{\linewidth}{!}{
\begin{tabular}{lccccccc}
    \hline
            Observable & $\qsq$ interval & SM &  \multicolumn{1}{c}{(\Cnine, \Cten)} &  \multicolumn{1}{c}{(\Cninep, \Ctenp)} & \multicolumn{1}{c}{(\Cnine, \Cnineee)} &  \multicolumn{1}{c}{(\ReCsevp, \ImCsevp)} &  \multicolumn{1}{c}{(\ReCsev, \ImCsev)}\\
    & & & $(-1.0, 0.0)$ & $(-0.2, -0.2)$ & $ (-1.0, 0.0)$ & $(0.00, 0.04)$ & $(-0.075, 0.000)$\\
        \hline
    $S_{3}(\BtoKstmumu)$ & $1.1  < \qsq < 2.5 \gev^2$ & $0.002$  &  -  & $0.011$ &  -  &  -  &  -  \\
    $S_{4}(\BtoKstmumu)$ & $1.1  < \qsq < 2.5 \gev^2$ & $-0.026$  & $-0.034$ & $-0.015$ & $-0.034$ &  -  &  -  \\
    $S_{5}(\BtoKstmumu)$ & $1.1  < \qsq < 2.5 \gev^2$ &  $0.056$ & $0.164$ & $0.035$ & $0.164$ &  -  &  -  \\
    $F_L(\BtoKstmumu)$ & $1.1  < \qsq < 2.5 \gev^2$ & $0.760$ & $0.682$ & $0.760$ & $0.682$ &  -  &  -   \\
    $A_\text{FB}(\BtoKstmumu)$ & $1.1  < \qsq < 2.5 \gev^2$ &  $-0.137$ & $-0.192$ & $-0.136$ & $-0.192$ &  -  &  -  \\
    $A_{7}(\BtoKstmumu)$ & $1.1  < \qsq < 2.5 \gev^2$ & $0.003$  & - & - & - &  $-0.037$  &  $0.003$  \\
    $A_{8}(\BtoKstmumu)$ & $1.1  < \qsq < 2.5 \gev^2$ &  $0.001$  & - & - & - &  $0.018$  &  $0.001$  \\
    $A_{9}(\BtoKstmumu)$ & $1.1  < \qsq < 2.5 \gev^2$ &  $0.000$  & - & - & - &  $-0.021$  &  $0.000$  \\
    $S_{3}(\BtoKstmumu)$ & $2.5  < \qsq < 4.0 \gev^2$ & $-0.011$  &  -  & $-0.002$ &  -  &  -  &  -  \\
    $S_{4}(\BtoKstmumu)$ & $2.5  < \qsq < 4.0 \gev^2$ & $-0.152$  & $-0.150$ & $-0.144$ & $-0.150$ &  -  &  -   \\
    $S_{5}(\BtoKstmumu)$ & $2.5  < \qsq < 4.0 \gev^2$ &  $-0.192$ & $-0.077$ & $-0.207$ & $-0.077$ &  -  &  -  \\
    $F_L(\BtoKstmumu)$ & $2.5  < \qsq < 4.0 \gev^2$ &  $0.796$ & $0.747$ & $0.797$ & $0.747$ &  -  &  -  \\
    $A_\text{FB}(\BtoKstmumu)$ & $2.5  < \qsq < 4.0 \gev^2$ & $-0.016$ & $-0.096$ & $-0.016$ & $-0.096$ &  -  &  -  \\
    $A_{7}(\BtoKstmumu)$ & $2.5  < \qsq < 4.0 \gev^2$ & $0.002$  & - & - & - &  $-0.026$  &  $0.002$  \\
    $A_{8}(\BtoKstmumu)$ & $2.5  < \qsq < 4.0 \gev^2$ & $0.001$  & - & - & - &  $0.015$  &  $0.001$  \\
    $A_{9}(\BtoKstmumu)$ & $2.5  < \qsq < 4.0 \gev^2$ & $0.000$  & - & - & - &  $-0.002$  &  $0.000$  \\
    $S_{3}(\BtoKstmumu)$ & $4.0  < \qsq < 6.0 \gev^2$ & $-0.025$  &  -  & $-0.017$ &  -  &  -   &  -  \\
    $S_{4}(\BtoKstmumu)$ & $4.0  < \qsq < 6.0 \gev^2$ &  $-0.225$ & $-0.221$ & $-0.219$ & $-0.221$ &  -   &  -  \\
    $S_{5}(\BtoKstmumu)$ & $4.0  < \qsq < 6.0 \gev^2$ & $-0.337$ & $-0.240$ & $-0.347$ & $-0.240$ &  -  &  -   \\
    $F_L(\BtoKstmumu)$ & $4.0  < \qsq < 6.0 \gev^2$ & $0.709$  & $0.687$ & $0.713$ & $0.687$ &  -  &  -  \\
    $A_\text{FB}(\BtoKstmumu)$ & $4.0  < \qsq < 6.0 \gev^2$ & $0.125$ & $0.041$ & $0.123$ & $0.041$ &  -  &  -   \\
    $A_{7}(\BtoKstmumu)$ & $4.0  < \qsq < 6.0 \gev^2$ & $0.002$  & - & - & - &  $-0.017$  &  $0.002$  \\
    $A_{8}(\BtoKstmumu)$ & $4.0  < \qsq < 6.0 \gev^2$ & $0.001$  & - & - & - &  $0.012$  &  $0.001$  \\
    $A_{9}(\BtoKstmumu)$ & $4.0  < \qsq < 6.0 \gev^2$ & $0.000$  & - & - & - &  $0.006$  &  $0.000$  \\
    $S_{3}(\BtoKstmumu)$ & $15  < \qsq < 19.0 \gev^2$ & $-0.204$ &  -  & $-0.203$ &  -  &  -  &  -   \\
    $S_{4}(\BtoKstmumu)$ & $15  < \qsq < 19.0 \gev^2$ &  $-0.301$ & $-0.301$ & $-0.301$ & $-0.301$ &  -  &  -   \\
    $S_{5}(\BtoKstmumu)$ & $15  < \qsq < 19.0 \gev^2$ & $-0.281$  & $-0.255$ & $-0.283$ & $-0.255$ &  -  &  -   \\
    $F_L(\BtoKstmumu)$ & $15.0  < \qsq < 19.0 \gev^2$ &  $0.343$ & $0.343$ & $0.343$ & $0.343$ &  -  &  -   \\
    $A_\text{FB}(\BtoKstmumu)$ & $15.0  < \qsq < 19.0 \gev^2$ &  $0.366$ & $0.329$ & $0.368$ & $0.329$ &  -  &  -   \\
    $A_{8}(\BtoKstmumu)$ & $15.0  < \qsq < 19.0 \gev^2$ & $0.000$  & - & - & - &  $0.003$  &  $0.000$  \\
    $A_{9}(\BtoKstmumu)$ & $15.0  < \qsq < 19.0 \gev^2$ &  $0.000$ & - & - & - &  $0.006$  &  $0.000$   \\
            \hline
  \end{tabular}}
  \end{center}
\label{tab:NPLHCb2}
\end{sidewaystable*}

\end{document}